\documentclass[prd,showpacs,preprintnumbers,aps,floats]{revtex4} 

\usepackage{graphicx}

\def\ltap{\ \raisebox{-.4ex}{\rlap{$\sim$}} \raisebox{.4ex}{$<$}\ }
\def\gtap{\ \raisebox{-.4ex}{\rlap{$\sim$}} \raisebox{.4ex}{$>$}\ }

\begin{document} 
\preprint{KIAS-03075}
\preprint{KEK-TH-859}
\preprint{YNU-HEPTh-03-103}
\preprint{TU-693}
\preprint{hep-ph/0311151}

\title{Beyond leading-order corrections to $\bar{B} \to X_s \gamma$ 
       at large $\tan \beta$: \\
       The charged-Higgs-boson contribution}
\author{Francesca~Borzumati$^{a}$, Christoph~Greub$^{b}$, 
          and Youichi~Yamada$^{c}$} 
\affiliation{%
${}^a$ KIAS, 207-43 Cheongryangri 2-dong, Dongdaemun-gu,      
       Seoul 130-722, Korea                                   \\
${}^b$ Institut f\"ur Theoretische Physik, Universit\"at
                Bern, Sidlerstrasse 5, 3012 Bern, Switzerland \\  
${}^c$ Department of Physics, Tohoku University,
                Sendai 980-8578, Japan}

\begin{abstract} 
Among the $O(\alpha_s\tan\!\beta)$ contributions to the Wilson
coefficients $C_7$ and $C_8$, relevant for the decay 
$\bar{B} \to X_s \gamma$, those induced by two-loop diagrams with
charged-Higgs-boson exchange and squark-gluino corrections are
calculated in supersymmetric models at large $\tan \beta$.  The
calculation of the corresponding Feynman integrals is exact, unlike in
previous studies that are valid when the typical supersymmetric scale
$M_{\rm SUSY}$ is sufficiently larger than the electroweak scale
$m_{\rm weak} (\sim m_W,m_t)$ and the mass of the charged Higgs boson
$m_{H^\pm}$.  Therefore, the results presented here can be used for
any value of the various supersymmetric masses. These results are
compared with those of an approximate calculation, already existing
in the literature, that is at the zeroth order in the expansion
parameter $(m_{\rm weak}^2,m_H^2)/M_{\rm SUSY}^2$, and with the
results of two new approximate calculations in which the first and
second order in the same expansion parameter are retained,
respectively. This comparison allows us to assess whether the results of
these three approximate calculations can be extended beyond the range
of validity for which they were derived, i.e., whether they can be
used for $m_H \gtap M_{\rm SUSY}$ and/or 
$M_{\rm SUSY} \sim m_{\rm weak}$.  It is found that the zeroth-order
approximation works well even for $m_H \gtap M_{\rm SUSY}$, provided 
$M_{\rm SUSY}^2 \gg m_{\rm weak}^2$. The inclusion of the higher-order
terms improves the zeroth-order approximation for 
$m_H^2 \ll M_{\rm SUSY}^2$, but it worsens it for 
$m_H \gtap M_{\rm SUSY}$.
\end{abstract} 

\vfill  

\maketitle

\setlength{\parskip}{1.01ex}

\section{Introduction}
\label{sect-intro}
While the rate for the inclusive decay $\bar{B} \to X_s \gamma$, 
${\rm BR}(\bar{B} \to X_s \gamma)$, has been calculated up to the
next-to-leading order (NLO) in QCD within the standard model
(SM)~\cite{NLOSM}, similarly precise calculations exist for only some
extensions of the SM.  Achieving a NLO precision for this decay rate
in such extensions, and in particular in supersymmetric models, is
important. It seems unlikely that an increased experimental precision
in the three apparatuses where this decay is measured, i.e. BABAR,
BELLE and CLEO~\cite{EXPERIMENTS}, combined with the increased
theoretical precision of a possible estimate of the next-to-NLO
corrections in the SM~\cite{NNLOproject}, will bring unequivocal
signals of new physics. Nevertheless, calculations of 
${\rm BR}(\bar{B} \to X_s \gamma)$ with NLO accuracy for the
models that are considered the most likely candidates to extend the
SM could help in understanding where the effective scale of these
models sets in, and the extent of the spreading of masses of
additional particles around this scale.

NLO calculations exist for two-Higgs-doublet models of type~I and
type~II~\cite{CRS,NLO2HDM-CDGG,NLO2HDM-BG}, as well as models in which
the couplings of the charged Higgs boson to fermions are, in absolute
values, those of type~II models, but are, in general,
complex~\cite{NLO2HDM-BG}.  In the case of supersymmetric models, the
situation is as follows.  For generic models, the QCD corrections to
the electroweak rate, calculated first in~\cite{BSGinSUSYproposal},
have been included only at leading-order (LO)
precision~\cite{LO-generalSUSY}. Higher-order QCD corrections have
been evaluated for specific scenarios.

In one class of such scenarios~\cite{NLO-SUSY1}, the two charginos,
one $\tilde{t}$ squark, which is predominantly right handed, and the
charged Higgs boson are assumed to be relatively light, while all the
other squarks and the gluino are heavy.  Moreover, no additional
sources of flavor violation are present at the electroweak scale,
other than the Cabibbo-Kobayashi-Maskawa mixing elements.

Other studies~\cite{NLO-SUSY2a,NLO-SUSY2b} considered a slightly
different class of scenarios, in which $\tan \beta$ is large, and the
supersymmetric spectrum is like the spectrum for the scenarios of
Ref.~\cite{NLO-SUSY1}, but without the assumption of a light
$\tilde{t}$ squark. The same minimality in flavor violation is also
assumed~\cite{MFVBuras}. The importance of supersymmetric models at
large $\tan \beta$ hardly needs to be highlighted here, as 
$\tan \beta$ tends to be large whenever an attempt to unify Yukawa
couplings is made, as required by a SO(10) grand
unification~\cite{largeTB}. Indeed, many phenomenological calculations
exists, in which the constraints imposed by the measured rate of
${\bar B} \to X_s \gamma$ to models at large $\tan \beta$ have been
analyzed at LO in QCD~\cite{bsglargeTB}.  The papers in
Refs.~\cite{NLO-SUSY2a,NLO-SUSY2b} allow one to refine these analyses 
in models which, as well as having large $\tan \beta$, predict the same 
type of supersymmetric spectrum that they assume.

The restriction to specific ranges of masses of the supersymmetric
partners of the SM particles, made in
Refs.~\cite{NLO-SUSY1,NLO-SUSY2a,NLO-SUSY2b}, has made possible the use
of an effective Lagrangian formalism, and has led to rather compact
and simple formulas. These very same restrictions, however, limit the
usefulness of such calculations, unless it is proven that the formulas
obtained in these papers can be safely used beyond the range of
validity for which they have been derived. Nevertheless, they have often
been employed also when the charged Higgs boson is as heavy as 
the squarks and the gluino. It is interesting to understand, in such
cases, how far the resulting analyses are from ideal ones, in 
which exact formulas are used.

\begin{figure}[h] 
\vspace{0.3truecm}
\begin{center} 
\includegraphics[width= 6cm]{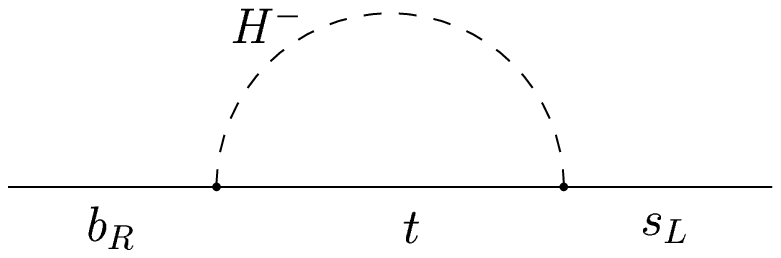}
\hspace{3mm}
\includegraphics[width= 6cm]{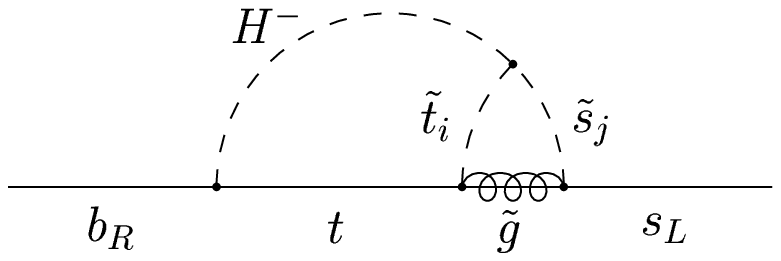}
\end{center} 
\vspace{-0.3truecm}
\caption[f1]{\small $b\to s\gamma$/$b\to s g$ by charged-Higgs-boson
 exchange, at one loop, on the left side, and with inclusion of
 gluino-squark subloops, on the right side. The photon/gluon is 
 assumed to be attached in all possible ways in both diagrams.}
\label{fig-CH-0and1}
\end{figure}

In this paper we address such an issue, focusing on the scenarios of
Refs.~\cite{NLO-SUSY2a,NLO-SUSY2b}. We consider the gluino-induced
supersymmetric corrections of $O(\alpha_s \tan \beta)$ that are the
largest beyond-leading-order corrections in scenarios with large 
$\tan \beta$.  Corrections of this type are obtained by (i)~including
the finite corrections to the $b$-quark mass in the
fermion--fermion--Higgs-boson and in the fermion--sfermion--Higgsino
couplings~\cite{dmb,carenaH0}, and~(ii) ``dressing'' with
squark-gluino subloops the one-loop diagrams mediated by the charged
Higgs and the charged Goldstone bosons that contribute to 
$\bar{B} \to X_s \gamma$ at the partonic level.  A graphical
representation of this ``dressing'' is explicitly shown in
Fig.~\ref{fig-CH-0and1}, for the diagrams contributing to 
$b\to s \gamma$ or to $b\to s g$ and mediated by the charged Higgs
boson.  (In this figure, the photon or the gluon is assumed to be
attached in all possible ways.)
Indeed, the substitution of the vertex $\bar{s}_L t_R H^-$ of the
one-loop diagrams with $\tilde{s}_L^\ast \tilde{t}_R H^-$, possible in
the two-loop diagrams, allows the gain of a $\tan \beta$ factor,
yielding the required $\alpha_s \tan \beta$.

As already pointed out, in the calculations of
Refs.~\cite{NLO-SUSY2a,NLO-SUSY2b}, all scalar superpartners of the SM
fermions, at the scale $M_{\rm SUSY}$, are assumed to be heavier than
the $t$ quark and the $W$ boson, whereas the charged Higgs boson
$H^\pm$ is assumed to be around the electroweak scale $m_{\rm weak}$,
with $m_{\rm weak} \sim M_W, m_t$. In the limit of large 
$M_{\rm SUSY}$, the two-loop diagrams in Fig.~\ref{fig-CH-0and1} in
which the photon is emitted only by the $t$ quark and the
charged Higgs boson, and the gluon only by the $t$ quark, all with
chirality flip on the $t$-quark line, are of nondecoupling nature. In
the same limit, all other diagrams in Fig.~\ref{fig-CH-0and1}
decouple.  It is conceivable, therefore, that for 
$m_{H^\pm}^2 \sim m_t^2 \ll M_{\rm SUSY}^2$, the former two-loop
diagrams are the only ones that give a sizable contribution to the
$O(\alpha_s \tan \beta)$ corrections to the decay amplitudes of
$b \to s \gamma$ and $b \to s g$, and that the expression for their
squark-gluino subloops is well approximated by the expression for the
same subdiagrams with vanishing external momenta.  Thus, in this 
approximation, these two-loop diagrams are factorized into two 
one-loop diagrams. Such a factorization is supported by the use of an
effective two-Higgs-doublet Lagrangian formalism, in which all heavy
degrees of freedom are integrated out. (A similar formalism was used
 in Ref.~\cite{DIAZ}, in which corrections of $O(\alpha \tan \beta)$ 
to the decay amplitude of $b\to s\gamma$ were calculated.)

A potential problem with this approach may arise when the
supersymmetric spectrum is not much heavier than $m_{\rm weak}$,
and/or when the mass of the charged Higgs boson tends to be closer to
$M_{\rm SUSY}$ than to $m_{\rm weak}$. (Papers with further
 improvements of the original calculations, such as those in
 Ref.~\cite{OTHERPAPS}, do not address this issue.)  One plausible
solution to this problem is to include in the original calculation 
higher-order terms, up to 
$O((m_{\rm weak}^2,m_{H^\pm}^2/M_{\rm SUSY}^2)^n)$, with a
suitable $n$. That is equivalent to saying that the effective
Lagrangian is extended to include higher-order operators. (A
discussion on this point can be found in Ref.~\cite{BGY}.)  An
efficient way to carry out this extension, consistently including all
the operators that yield terms of the same order $n$, is to make use
of the heavy mass expansion (HME)~\cite{HME}. Starting, for example,
from the regime in which 
$m_t^2, M_W^2 \sim m_{H^\pm}^2 \ll M_{\rm SUSY}^2$, this technique 
allows one to add all operators needed when $m_{H^\pm}$
tends to $M_{\rm SUSY}$ from below. It remains, however, to be
established up to what value of $n$ it is necessary to extend this
expansion, in order to obtain an estimate of 
${\rm BR}(\bar{B}\to X_s \gamma)$ adequate for all possible values of
$M_{\rm SUSY}$ and of $m_{H^\pm}$ predicted by different
supersymmetric models at large $\tan \beta$.  Clearly, an accurate
answer to this question can be given only by a comparison with the
exact calculation of all the two-loop diagrams that give rise to
$O(\alpha_s \tan \beta)$ corrections, in which no assumption is
made on the relative size of $m_{H^\pm}$, $M_{\rm SUSY}$, and 
$m_{\rm weak}$.

In this paper we perform such an exact calculation of these two-loop
diagrams contributing to the two Wilson coefficients $C_7$ and $C_8$
related to the partonic processes $b \to s \gamma$ and $b \to s g$.
Among all the contributions to these
processes~\cite{BSGcontributions}, we restrict ourselves to those
mediated by the charged Higgs boson, addressing first the problem of
obtaining results valid throughout all values of $m_{H^\pm}$. We
postpone the presentation of the complete
${\rm BR}(\bar{B}\to X_s \gamma)$ to future work~\cite{BGYfuture}.

We also calculate the same diagrams using the HME technique, up to
order $O((m_{\rm weak}^2,m_{H^\pm}^2/M_{\rm SUSY}^2)^2)$. The
hope is that, in the case in which the series expansion appears to be
quickly converging, approximate and possibly still compact formulas
can be provided, which nevertheless allow one to extend the results
existing in the literature to all possible values of $m_{H^\pm}$ and
$M_{\rm SUSY}$. It turns out that the series expansion in
$(m_{\rm weak}^2,m_{H^\pm}^2/M_{\rm SUSY}^2)$ is far from being
as well behaved as hoped and that, in general, the use of the exact
results cannot be avoided.  However, whereas the approximate 
calculations including terms up to
$O((m_{\rm weak}^2,m_{H^\pm}^2/M_{\rm SUSY}^2)^2)$ give results
in disagreement with the exact one for values of
$m_{H^\pm} \gtap M_{\rm SUSY}$, the lowest-order term of the series
expansion, i.e., the nondecoupling approximation of 
Refs.~\cite{NLO-SUSY2a,NLO-SUSY2b}, seems to be a good approximation
of the exact result throughout all ranges of $m_{H^\pm}$, provided
$M_{\rm SUSY}$ is sufficiently larger than $m_{\rm weak}$. Deviations
appear for supersymmetric particles not much above the electroweak
scale.

The paper is organized as follows. In Sec.~\ref{sect-diagrams}, we
list all the diagrams needed to evaluate the charged-Higgs-boson 
mediated $O(\alpha_s \tan \beta)$ contributions to the Wilson
coefficients $C_7$ and $C_8$ entering in the calculation of the 
${\rm BR}(\bar{B}\to X_s \gamma)$.  We also give these coefficients at
the matching electroweak scale $\mu_W$ in terms of two-loop scalar
integrals.  In Sec.~\ref{sect-EFLvsHME} we review some basic
properties of the HME technique and discuss how it can be applied to
our specific case.  In Sec.~\ref{sect-WCoeff} we show numerical
results for the charged-Higgs-boson contributions to $C_7(\mu_W)$ and
$C_8(\mu_W)$, obtained from the exact two-loop calculation, and from 
the two approximate calculations including up to the first two orders 
in $(m_{\rm weak}^2,m_{H^\pm}^2)/M_{\rm SUSY}^2$. These results are
compared and discussed.  Finally, in Sec.~\ref{sect-conclu} we 
conclude.

\section{The two-loop diagrams}
\label{sect-diagrams}
 
\begin{figure}[t] 
\begin{center} 
\includegraphics[width= 6cm]{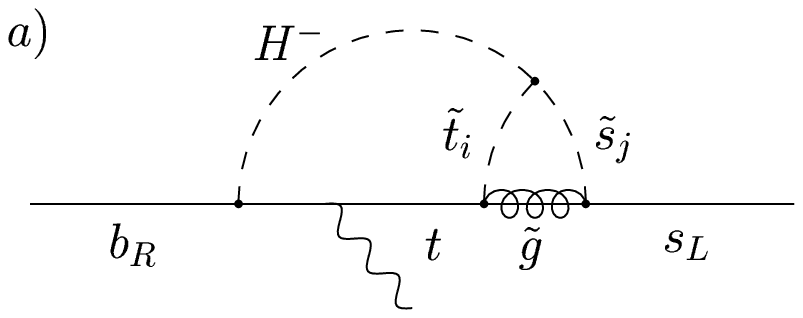}
\hspace{3mm}
\includegraphics[width= 6cm]{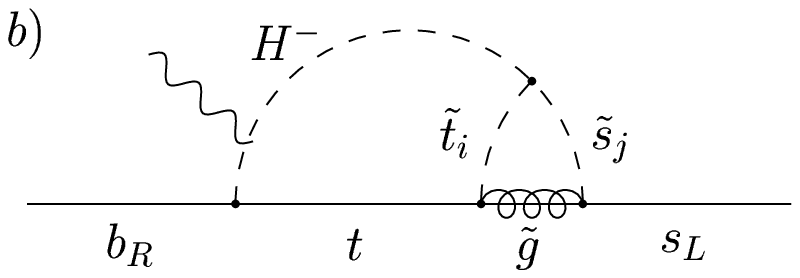}
\end{center} 
\vspace{-0.1truecm}
\begin{center} 
\includegraphics[width= 6cm]{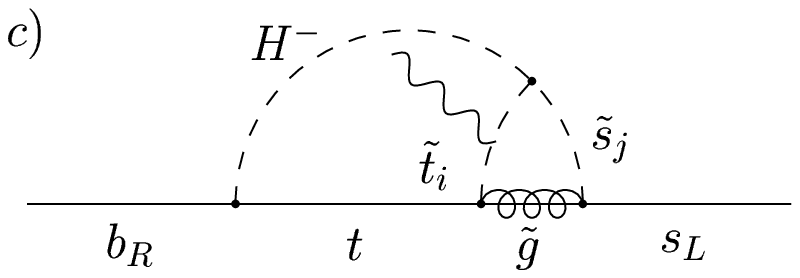}
\hspace{3mm}
\includegraphics[width= 6cm]{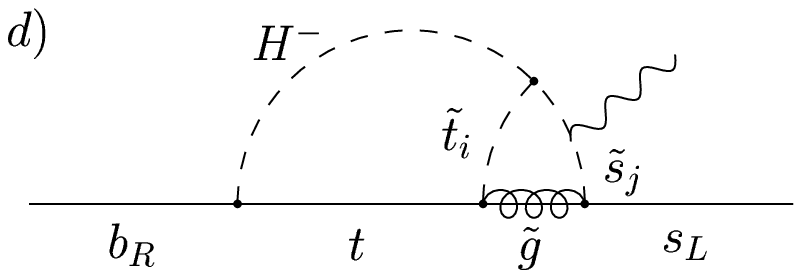}
\end{center} 
\vspace{-0.2truecm}
\begin{center} 
\includegraphics[width= 6cm]{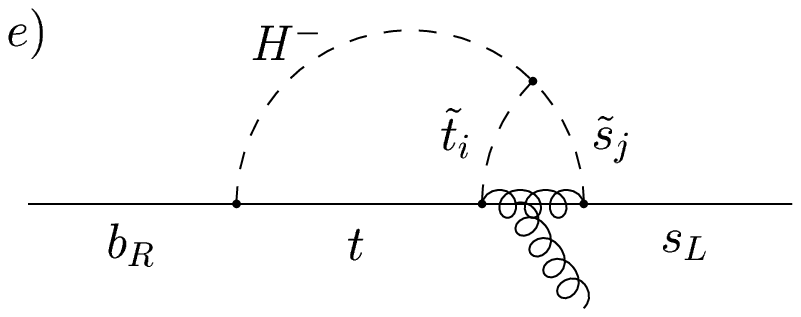}
\end{center} 
\vspace{-0.3truecm}
\caption[f1]{\small Charged-Higgs-boson mediated diagrams contributing 
 at order $O(\alpha_s \tan \beta)$ to the partonic decays 
 $b\to s\gamma$ and $b\to s g$. The photon must be replaced by a 
 gluon, and vice versa, whenever possible.} 
\label{fig-CHexch2loops}
\end{figure}

The charged-Higgs-boson mediated two-loop diagrams, contributing at 
$O(\alpha_s \tan \beta)$ to the partonic decays 
$b\to s \gamma$ and $b \to s g$, are obtained from the right diagram
of Fig.~\ref{fig-CH-0and1} after allowing a photon and/or a gluon to
be emitted in all possible ways.  They are shown explicitly in
Fig.~\ref{fig-CHexch2loops}, where the photon may be replaced by a
gluon, and vice versa, whenever possible.

It should be observed that, while the one-loop diagram on the left
side of Fig.~\ref{fig-CH-0and1} has a chirality flip in the internal
$t$-quark line (we neglect in this discussion the diagram with
chirality flip on the external $b$-quark line, which is $\tan \beta$
suppressed), the diagrams in Fig.~\ref{fig-CHexch2loops} can have such
a chirality flip on the $\tilde{t}$-squark line also.  The number of
contributions to be evaluated, therefore, amounts to 8 for the
calculation of the Wilson coefficient $C_7(\mu_W)$, and 8 for the
coefficient $C_8(\mu_W)$.  The diagrams used for the calculations of
${\rm BR}(\bar{B}\to X_s \gamma)$ in
Refs.~\cite{NLO-SUSY2a,NLO-SUSY2b} are the two at the top of
Fig.~\ref{fig-CHexch2loops} for the calculation of $C_7(\mu_W)$,
i.e.,~\ref{fig-CHexch2loops}(a) and~\ref{fig-CHexch2loops}(b), and the
diagram~\ref{fig-CHexch2loops}(a), with the photon replaced by a gluon,
for the calculation of $C_8(\mu_W)$, all with chirality flip on the
$t$-quark line only.

We have calculated all the 16 contributing terms, by making use of
results and techniques presented in Ref.~\cite{GvdB}.  Our
normalization of the Wilson coefficients $C_7(\mu_W)$ and $C_8(\mu_W)$
is the conventional one, as follows from the definition of the
effective Hamiltonian,
\begin{equation}
H_{\rm eff} \supset
 -\frac{4G_F}{\sqrt{2}}V^*_{ts}V_{tb}
 \left[ C_7(\mu) O_7(\mu) + C_8(\mu) O_8(\mu) 
 \right],
\label{eq-EffHam}
\end{equation}
and of the operators $O_7$ and $O_8$, 
\begin{eqnarray}
O_7(\mu) = 
 \frac{e}{16\pi^2} m_b(\mu)\bar{s}_L\sigma^{\mu\nu} b_R F_{\mu\nu}, 
\hspace{0.8truecm}
O_8(\mu) = 
 \frac{g_s}{16\pi^2}m_b(\mu)\bar{s}_L\sigma^{\mu\nu}T^a b_RG^a_{\mu\nu}, 
\label{eq-O7and8}
\end{eqnarray}
where $F_{\mu\nu}$ and $G^a_{\mu\nu}$ are the field strengths of the
photon and the gluon, respectively.  We denote by 
$ C_{7,H}(\mu_W)$ and $ C_{8,H}(\mu_W)$ the $\tan \beta$-unsuppressed
charged-Higgs-boson contribution to $C_7(\mu_W)$ and
$C_8(\mu_W)$, and we decompose them as
\begin{equation}
C_{i,H}(\mu_W) = \frac{1}{1+\Delta_{b_R,b} \tan\beta}
\left[ C_{i,H}^0(\mu_W) + \Delta C_{i,H}^1(\mu_W) \right],
\label{defWC}
\end{equation}
where $C_{i,H}^0(\mu_W)$ and $\Delta C_{i,H}^1(\mu_W)$ are induced by
the one-loop diagram in Fig.~\ref{fig-CH-0and1} and the two-loop
diagrams in Fig.~\ref{fig-CHexch2loops}, respectively.  The overall
factor $1/(1+\Delta_{b_R,b} \tan\beta)$ (see notation of
Ref.~\cite{BGY}) stems from expressing the $H^+\bar{t}_L b_R$ Yukawa
coupling in terms of $m_b$, corrected up to 
$O(\alpha_s \tan \beta)$~\cite{dmb,carenaH0}.  $\Delta_{b_R,b}$ is 
given by
\begin{equation}
\Delta_{b_R,b} = 
\frac{C_F \alpha_s}{2\pi} \mu m_{\tilde{g}} \,
 I(m_{\tilde{b}_1}^2,m_{\tilde{b}_2}^2,m_{\tilde{g}}^2),
\label{deltamb}
\end{equation}
where the function $I$ is
\begin{widetext}
\begin{equation}
I(m_1^2,m_2^2,m_3^2)  =\, - 
 \frac{
     m_1^2 m_2^2  \ln(m_1^2/m_2^2) 
+    m_2^2 m_3^2  \ln(m_2^2/m_3^2) 
+    m_3^2 m_1^2  \ln(m_3^2/m_1^2) }
      { 
    (m_1^2-\!m_2^2)
    (m_2^2-\!m_3^2)
    (m_3^2-\!m_1^2) }.
\label{c0function}
\end{equation}
\end{widetext}

As the photon can be emitted by the $t$ quark and the
charged Higgs boson, we have two contributions to $C_7^0(\mu_W)$:
\begin{eqnarray}
 C_{7,H}^{0\,a}(\mu_W) & = & 
  -\frac{1}{2} \, Q_t \, \frac{m_t^2}{m_{H^\pm}^2} 
 F_3\left(\!\!\frac{m_t^2}{m_{H^\pm}^2} \!\!\right),  
\nonumber \\
 C_{7,H}^{0\,b}(\mu_W) & = & 
  \frac{1}{2} \, Q_H \, \frac{m_t^2}{m_{H^\pm}^2} 
 F_4\left(\!\!\frac{m_t^2}{m_{H^\pm}^2} \!\!\right),  
\label{eq-c70}
\end{eqnarray}
while only one to $C_8^0(\mu_W)$:
\begin{equation}
 C_{8,H}^{0\,a}(\mu_W) \ = \ 
  -\frac{1}{2} \frac{m_t^2}{m_{H^\pm}^2} 
 F_3\left(\!\!\frac{m_t^2}{m_{H^\pm}^2} \!\!\right),   
\label{eq-c80}
\end{equation}
due to the emission of the gluon from the $t$ quark.  The superscript
indices $a$ and $b$ in the above expressions denote emission of the
photon/gluon from the $t$ quark and the charged Higgs boson,
respectively.  The functions $F_3$ and $F_4$ are listed in
Appendix~\ref{app-funct}.  $Q_t$ and $Q_H$ indicate the electric
charges of the $t$ quark and the charged Higgs boson ($Q_H=-1$).

The contributions to $\Delta C_{7,H}^1(\mu_W)$, due to the 
different two-loop diagrams in Fig.~\ref{fig-CHexch2loops}, are
\begin{eqnarray}
\Delta C_{7,H}^{1\,a}(\mu_W) 
& = & 
\frac{1}{2} \,Q_t \,C_F 
\left(\alpha_s \tan\beta\right)  m_t \,\mu \ (4\pi)^3
\left[ (U_t)_{2i} (U^T_t)_{i2} \,m_t m_{\tilde{g}} \,I_{ti2} 
      -(U_t)_{2i} (U^T_t)_{i1} I_{ti1} \right],
\nonumber \\
\Delta C_{7,H}^{1\,b}(\mu_W) 
& = & 
\frac{1}{2} \,Q_H C_F 
\left(\alpha_s  \tan\beta\right)  m_t \,\mu \ (4\pi)^3
\left[ (U_t)_{2i} (U^T_t)_{i2} \,m_t m_{\tilde{g}} \,I_{Hi2}  
      -(U_t)_{2i} (U^T_t)_{i1} I_{Hi1} \right],
\nonumber \\
\Delta C_{7,H}^{1\,c}(\mu_W)  
& = & 
\frac{1}{2} \,Q_{\tilde{t}} \,C_F 
\left(\alpha_s \tan\beta\right)  m_t \,\mu \ (4\pi)^3
\left[ (U_t)_{2i} (U^T_t)_{i2} \,m_t m_{\tilde{g}} \,I_{\tilde{t}i2} 
      -(U_t)_{2i} (U^T_t)_{i1} I_{\tilde{t}i1} \right],  
\nonumber \\
\Delta C_{7,H}^{1\,d}(\mu_W)  
& = & 
\frac{1}{2} \,Q_{\tilde{s}} \,C_F 
\left(\alpha_s \tan\beta\right)  m_t \,\mu \ (4\pi)^3
\left[ (U_t)_{2i} (U^T_t)_{i2} \,m_t m_{\tilde{g}} \,I_{\tilde{s}i2} 
      -(U_t)_{2i} (U^T_t)_{i1} I_{\tilde{s}i1} \right].  
\label{eq-deltaC7}
\end{eqnarray}
In the above expressions, the definitions 
\begin{equation}
 \tan \beta = \frac{v_U}{v_D},
\quad \quad \quad 
 \bar{v}^2  =  \frac{v_U^2 + v_D^2}{2}
\label{eq-vevsdef}
\end{equation}
are adopted, with $v_U$ and $v_D$ the vacuum expectation values of the
neutral components of the Higgs doublet with hypercharge $+1/2$ and
$-1/2$. All phases for the supersymmetric parameters are assumed to be
vanishing.  Thus, the matrix $U_t$ is the $2\times 2$ orthogonal
diagonalization matrix of the matrix for the $\tilde{t}$-squark mass
squared, and a summation over the index $i$, identifying the two
$\tilde{t}$ eigenvalues, $\tilde{t}_1$ and $\tilde{t}_2$, is
understood.  Our conventions for the squark sectors are listed in
Appendix~\ref{app-conventions}.  We have also assumed that the
left-right mixing in the matrix for the $\tilde{s}$-squark mass
squared is nearly vanishing and therefore only one $\tilde{s}$-squark
eigenstate, the left-handed one, denoted simply by $\tilde{s}$, is
needed for this calculation.  $Q_{\tilde{t}}$ and $Q_{\tilde{s}}$
indicate the electric charges of the $\tilde{t}$ and the
$\tilde{s}$ squarks.  $C_F=4/3$ is a color factor.  Finally, the
scalar integrals $I_{ti1}$, $I_{ti2}$, $I_{Hi1}$, $I_{Hi2}$,
$I_{\tilde{t}i1}$, $I_{\tilde{t}i2}$, $I_{\tilde{s}i1}$, and
$I_{\tilde{s}i2}$ are listed in integral form in 
Appendix~\ref{app-integrals}.

Similarly, the contributions to $\Delta C_{8,H}^1(\mu_W)$ are
\begin{eqnarray}
\Delta C_{8,H}^{1\,a}(\mu_W) 
& = & 
\frac{1}{2}\, C_F \,
\left(\alpha_s \tan\beta\right)  m_t \,\mu \ (4\pi)^3
\left[ (U_t)_{2i} (U^T_t)_{i2} \,m_t m_{\tilde{g}} \, I_{ti2} 
      -(U_t)_{2i} (U^T_t)_{i1} I_{ti1} \right],
\nonumber \\
\Delta C_{8,H}^{1\,c}(\mu_W)  
& = & 
\frac{1}{2} \,C_F \left(1\!-\! \frac{C_V}{2 C_F}\right) 
\left(\alpha_s \tan\beta\right)  m_t \,\mu \ (4\pi)^3
\left[ (U_t)_{2i} (U^T_t)_{i2} \,m_t m_{\tilde{g}} \,I_{\tilde{t}i2} 
      -(U_t)_{2i} (U^T_t)_{i1} I_{\tilde{t}i1} \right],  
\nonumber \\
\Delta C_{8,H}^{1\,d}(\mu_W)  
& = & 
\frac{1}{2} \,C_F \left(1\!-\! \frac{C_V}{2 C_F}\right) 
\left(\alpha_s \tan\beta\right)  m_t \,\mu \ (4\pi)^3
\left[ (U_t)_{2i} (U^T_t)_{i2} \,m_t m_{\tilde{g}} \,I_{\tilde{s}i2} 
      -(U_t)_{2i} (U^T_t)_{i1} I_{\tilde{s}i1} \right],  
\nonumber \\
\Delta C_{8,H}^{1\,e}(\mu_W) 
& = & 
\frac{1}{2} \, C_V \,
\left(\alpha_s \tan\beta \right) m_t \,\mu \ (4\pi)^3
\left[ (U_t)_{2i} (U^T_t)_{i2} \,m_t m_{\tilde{g}} \,I_{\tilde{g}i2} 
      -(U_t)_{2i} (U^T_t)_{i1} I_{\tilde{g}i1} \right],  
\label{eq-deltaC8}
\end{eqnarray}
where, obviously, $\Delta C_{8,H}^{1\,b}(\mu_W)=0$. Here, $C_V=3$ is
another SU(3) group factor and the integrals 
$I_{\tilde{g}i1}$,$I_{\tilde{g}i2}$ are also defined in
Appendix~\ref{app-integrals}. The explicit expression of the integrals
introduced in this section can be obtained upon request as a FORTRAN
code.

The results shown in Refs.~\cite{NLO-SUSY2a,NLO-SUSY2b} for the 
${\rm BR}(\bar{B}\to X_s \gamma)$ are obtained using only the first 
term in the square brackets of $\Delta C_{7,H}^{1\,a}(\mu_W)$, 
$\Delta C_{7,H}^{1\,b}(\mu_W)$, and $\Delta C_{8,H}^{1\,a}(\mu_W)$, 
with the approximations for $I_{ti2}$ and $I_{Hi2}$ listed in 
Appendix~\ref{app-integrals}.  As anticipated in the Introduction, we
call the approximation of these references the nondecoupling
approximation. It is indeed of nondecoupling type in the limit of
heavy supersymmetric particles and collects all terms of
$O((m_{\rm weak}^2,m_{H^\pm}^2/M_{\rm SUSY}^2)^0)$, as explained in
greater details in the next section.  Strictly speaking, however, it
contains formally decoupling terms coming from the masses and the
couplings of squarks (see Ref.~\cite{OTHERPAPS}).

\section{Effective Lagrangian and Heavy Mass Expansion}
\label{sect-EFLvsHME}
The calculation of the two-loop diagrams discussed in the previous
section was performed in Refs.~\cite{NLO-SUSY2a,NLO-SUSY2b} under 
the assumption of heavy squarks and gluino, at the scale $M_{\rm SUSY}$,
and a light charged Higgs boson, of $O(m_{\rm weak})$ 
($m_{\rm weak}\sim M_W, m_t$), with 
$m_{\rm weak}^2 \ll M_{\rm SUSY}^2$. In these calculations, terms of 
$O((m_{\rm weak}^2,m_{H^\pm}^2/M_{\rm SUSY}^2)^n)$ with 
$n>0$, i.e., terms that decouple in the limit $M_{\rm SUSY}\to \infty$,
were neglected except for those trivially accounted for by the couplings
of squarks and their masses~\cite{OTHERPAPS}.

The amplitude of the diagrams in Fig.~\ref{fig-CHexch2loops} can be
expanded in $(m_{\rm weak}^2,m_{H^\pm}^2/M_{\rm SUSY}^2)$. It seems
plausible that by retaining in such an expansion terms up to
$O((m_{\rm weak}^2,m_{H^\pm}^2/M_{\rm SUSY}^2)^n)$ with 
$n\ge 1$, may allow one to extend the results obtained to values of
$m_{H^\pm}$ not too dissimilar from $M_{\rm SUSY}$, as 
well as to a not too large value for $M_{\rm SUSY}$.  In this spirit,
it seems worth performing these approximate calculations, in the hope
that the results turn out to be still relatively compact.

A systematic way to include all the needed terms at 
$O((m_{\rm weak}^2,m_{H^\pm}^2/M_{\rm SUSY}^2)^n)$ is provided by 
the heavy mass expansion technique~\cite{HME}.  Before proceeding
further, we recall here some basic properties of this expansion.

Let us assume that all masses of a given Feynman diagram $\Gamma$ can
be divided into a set of large
\mbox{$\underline{M}=\{M_{1},M_{2},\ldots$\}} and small
$\underline{m}=\{m_{1},m_{2}, \ldots\}$ masses. If all external
momenta $\underline{q}=\{q_{1},q_{2},\ldots\}$ are small compared to
the scale of the large masses $\underline M$, then the dimensionally
regularized (unrenormalized) Feynman integral $F_{\Gamma}$ associated
with the Feynman diagram $\Gamma$ can be decomposed as
\begin{equation}
F_{\Gamma} \stackrel{\underline{M} \to \infty}{\sim}
 \sum_{\gamma} F_{\Gamma / \gamma} \circ
         {\cal T}_{\underline{q}^{\gamma},\underline{m}^{\gamma}} 
   F_{\gamma}\left(\underline{q}^{\gamma},\underline{m}^{\gamma},
                   \underline{M}\right),
\label{eq-HMEbasic}
\end{equation}
where the sum is performed over all subdiagrams $\gamma$ of $\Gamma$
which (i) contain all lines with heavy masses ($\underline{M}$),
and (ii) are one-particle irreducible with respect to lines with
small masses ($\underline{m}$). The case $\gamma=\Gamma$ is always
included in the sum $\sum_{\gamma}$. For each $\gamma$,
$\underline{m}^{\gamma}$ denotes the set of light masses,
$\underline{q}^{\gamma}$ the set of all external momenta with respect
to the subdiagram $\gamma$, which can be internal momenta with respect
to the full diagram $\Gamma$.  The operator ${\cal T}$ performs a
Taylor expansion in the variables $\underline{q}^{\gamma}$ and
$\underline{m}^{\gamma}$ and it is understood to act directly on the
integrand of the subdiagram $\gamma$.  The diagram $\Gamma/\gamma$ is
obtained by reducing $\gamma$ to a vertex in $\Gamma$.  Thus, by
factorizing the product of scalar propagators of the original, say,
$l$-loop diagram $\Gamma$ as 
$\Pi_{\Gamma} = \Pi_{\Gamma / \gamma}\Pi_{\gamma}$, the decomposition
of the original Feynman integral $F_{\Gamma}$ is simply
\begin{equation}
F_{\Gamma / \gamma} \circ  {\cal T}_{\underline{q}^{\gamma},
 \underline{m}^{\gamma}}       F_{\gamma} 
=
\int dk_{1} \cdots  dk_{l} \, \Pi_{\Gamma / \gamma} 
\, {\cal T}_{\underline{q}^{\gamma},
\underline{m}^{\gamma}}   \Pi_{\gamma}.
\label{eq-factor}
\end{equation}
Note that the Taylor operator ${\cal T}$ introduces additional
spurious IR or UV divergences in the various terms of the
sum~$\sum_{\gamma}$, which cancel in the sum.

\begin{figure}[t] 
\begin{center} 
\includegraphics[width= 6.3cm]{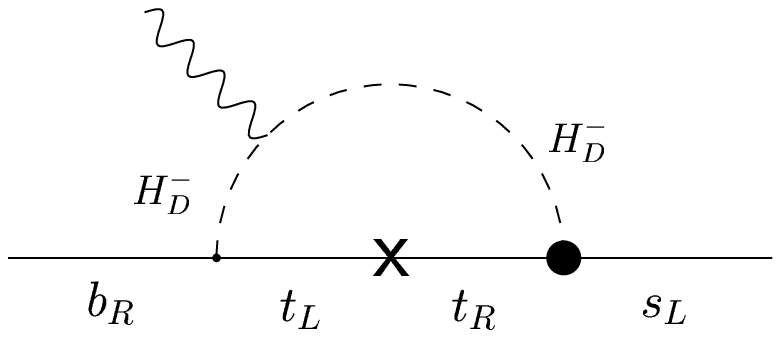}
\hspace*{1.5truecm}
\includegraphics[width= 4.0cm]{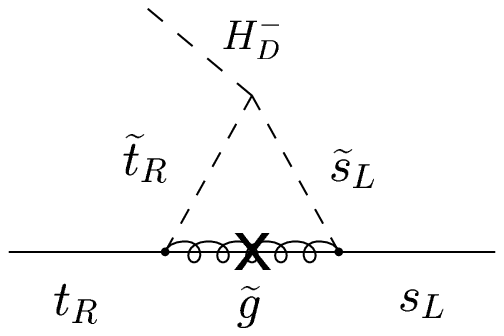}
\end{center} 
\begin{center} 
\includegraphics[width= 6.3cm]{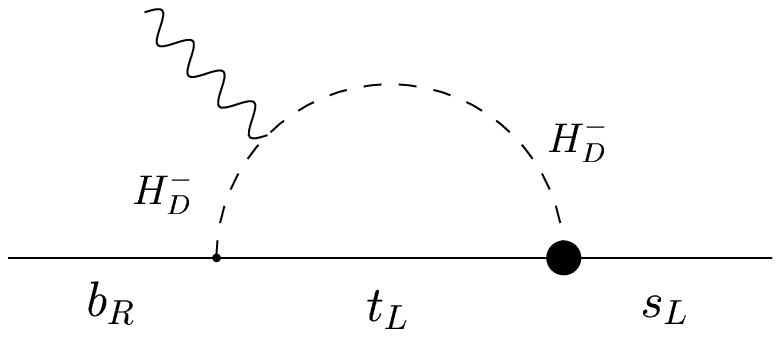}
\hspace*{1.5truecm}
\includegraphics[width= 4.0cm]{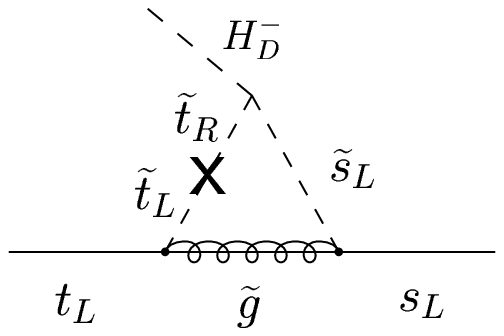}
\end{center} 
\vspace{-0.3truecm}
\caption[f1]{\small The diagrams $\Gamma/\gamma_1$, on the left side,
 obtained from the diagram (b) in Fig.~\ref{fig-CHexch2loops}. The
 corresponding $\gamma_1$ are shown on the right side. [See 
 Eq.~(\ref{eq-HMEbasic}) in the text.] The  chirality flip is on the
 $t$-quark line in the upper two diagrams, on the $\tilde{t}$-squark
 line on the lower two. All diagrams in this figure and in
 Figs.~\ref{fig-boxphsubdiagr} and~\ref{fig-boxglsubdiag} below are
 shown in the gauge eigenbasis for squarks and Higgs bosons.} 
\label{fig-triansubdiag} 
\end{figure}

\begin{figure}[t] 
\begin{center} 
\includegraphics[width= 6.3cm]{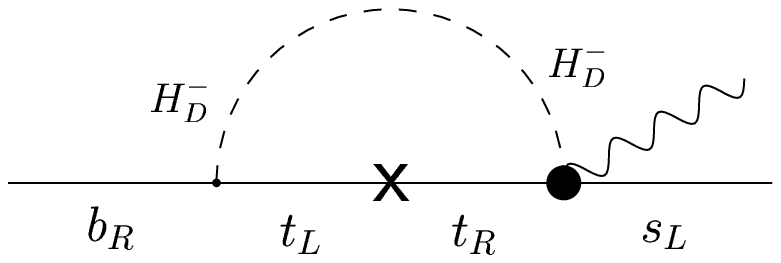}
\hspace*{1.5truecm} 
\includegraphics[width= 4.0 cm]{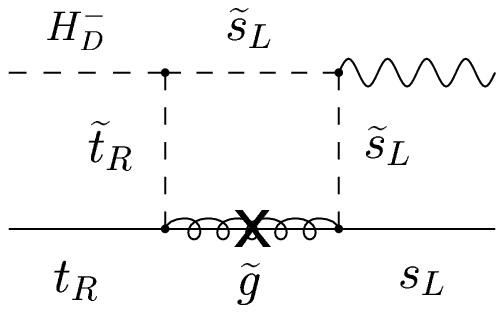}
\end{center} 
\begin{center} 
\includegraphics[width= 6.3cm]{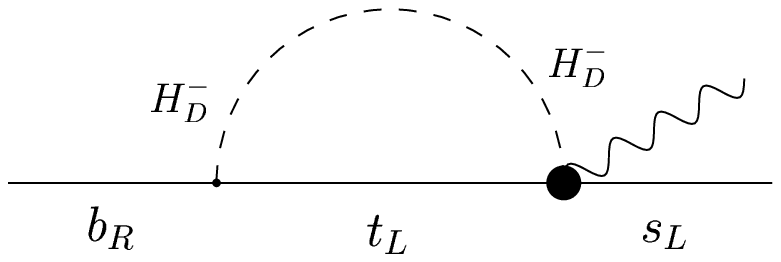}
\hspace*{1.5truecm} 
\includegraphics[width= 4.0 cm]{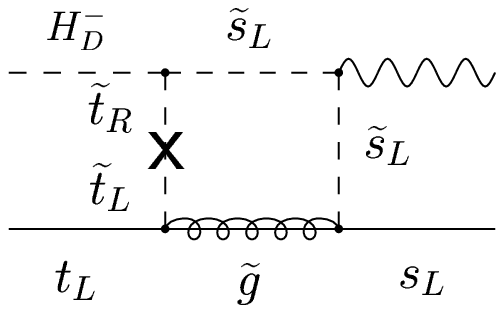}
\end{center}
\vspace{-0.3truecm} 
\caption[f1]{\small Same as in Fig.~\ref{fig-triansubdiag}, but with the  
 photon radiated off the $\tilde{s}_L$ line.} 
\label{fig-boxphsubdiagr} 
\begin{center} 
\includegraphics[width= 6.3cm]{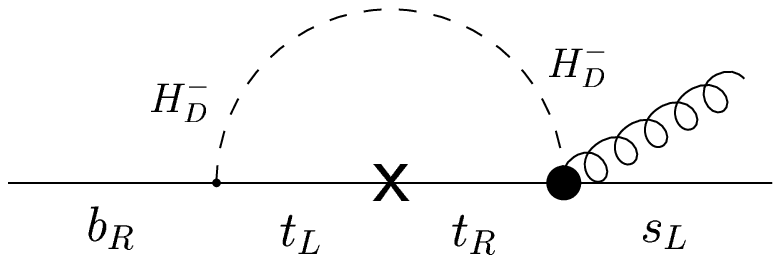}
\hspace*{1.5truecm} 
\includegraphics[width= 4.0 cm]{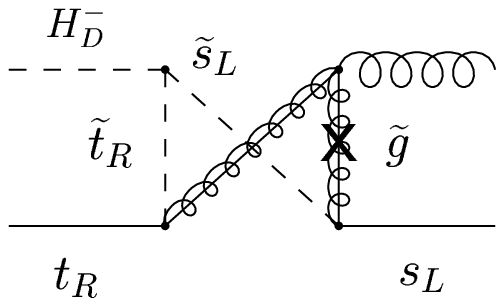}
\end{center} 
\begin{center} 
\includegraphics[width= 6.3cm]{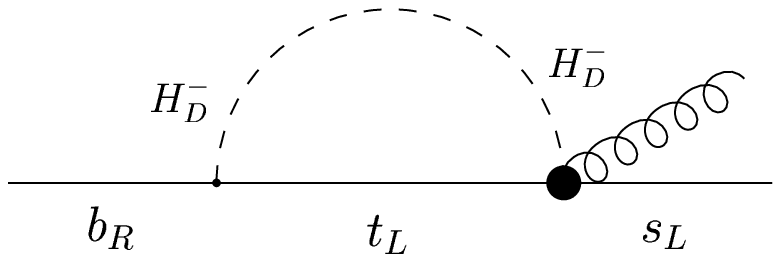}
\hspace*{1.5truecm} 
\includegraphics[width= 4.0 cm]{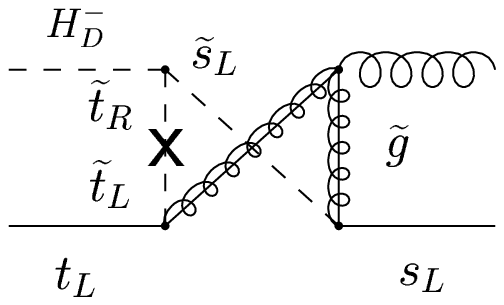}
\end{center}
\vspace{-0.3truecm} 
\caption[f1]{\small The diagrams $\Gamma/\gamma_1$ for the transition
 $b \to s g$, on the left side, with the gluon emitted by the gluino
 $\tilde{g}$. The corresponding $\gamma_1$ are shown on the right side.
 The chirality flip is on the $t$-quark line in the upper two diagrams and
 on the $\tilde{t}$-squark line on the lower two.} 
\label{fig-boxglsubdiag} 
\end{figure}

For the calculation of the diagrams in Fig.~\ref{fig-CHexch2loops},
we start assuming that 
\begin{equation}
m_{H^\pm}^2 \sim m_{\rm weak}^2 \ll  M_{\rm SUSY}^2, 
\label{eq-decouplimit}
\end{equation}
where $m_{\rm weak}$ is, in turn, $\gg m_b$.  In this limit, the
diagram (b) in Fig.~\ref{fig-CHexch2loops} with chirality flip on the
$t$-quark line has a Feynman integral that can be decomposed into two
terms contributing to the sum $\sum_\gamma$ of
Eq.~(\ref{eq-HMEbasic}). In the first term, $\gamma_1$ is the upper
right diagram of Fig.~\ref{fig-triansubdiag}, and the corresponding
Feynman integral must be expanded in the momenta of the external
particles, of the charged Higgs boson ${H^\pm}$, and of the $t$ and
$s$ quarks; $\Gamma/\gamma_1$ is the diagram on the upper
left of the same figure.  Although the calculation is actually done 
in the mass eigenbasis for squarks and Higgs bosons, the diagrams in 
this figure (as well as those in Figs.~\ref{fig-boxphsubdiagr} 
and~\ref{fig-boxglsubdiag}) are shown in the gauge eigenbasis. In an 
effective two-Higgs-doublet Lagrangian, obtained after integrating out
all heavy degrees of freedom, i.e.,  squarks and gluino, the zeroth-order
expansion of $\gamma_1$ is described by the term in the Lagrangian
\begin{equation}
{\cal L}^{\rm eff} 
 \ = \  - 
 V_{ts} \frac{m_t({\rm SM})}{\bar{v}\sin \beta}
  \Delta_{t_R,s} \,
 H_D^+ \bar{t}_R s_L + ({\rm H.c.}),  
\label{eq-2HDMEffLag}
\end{equation}
where, again, we follow the notation of Ref.~\cite{BGY}, and the
coefficient $\Delta_{t_R,s}$ is
\begin{equation}
\Delta_{t_R,s} = \frac{C_F \alpha_s}{2\pi} \mu m_{\tilde{g}} 
(U_t)_{2i} (U^T_t)_{i2} \,
I(m_{\tilde{t}_i}^2,m_{\tilde{s}}^2,m_{\tilde{g}}^2) \,.
\label{eq-effvertex}
\end{equation}
The function $I$ is the same as in Eq.~(\ref{c0function}). This zeroth 
order corresponds to the nondecoupling approximation of 
Refs.~\cite{NLO-SUSY2a,NLO-SUSY2b} and gives rise to the following 
contribution to the Wilson coefficient $C_{7,H}(\mu_W)$:
\begin{eqnarray} 
\Delta C_{7,H}^{1\,b}(\mu_W)\vert_{\rm nondec}   
& = &  
-\frac{1}{2} \,Q_H \,C_F 
\left(\alpha_s \tan\beta\right) \, \mu m_{\tilde{g}} \ \frac{1}{2\pi}
 (U_t)_{2i} (U^T_t)_{i2} \ 
 I(m_{\tilde{t}_i}^2,m_{\tilde{s}}^2,m_{\tilde{g}}^2) \,   
 \frac{m_t^2}{M_{H^\pm}^2} 
 F_4 \!\left(\!\!\frac{m_t^2}{M_{H^\pm}^2}\!\!\right) 
\nonumber \\  
& = & 
-\frac{1}{2} \,Q_H \, \tan\beta \, \Delta_{t_R,s} \, 
 \frac{m_t^2}{M_{H^\pm}^2} 
F_4 \!\left(\!\!\frac{m_t^2}{M_{H^\pm}^2}\!\!\right). 
\label{C7Hnondec}
\end{eqnarray}
The second term contributing to the sum~$\sum_\gamma$, $\gamma_2$, is
the full diagram, and its Feynman integral is to be expanded in 
$m_t$,$m_{H^\pm}$,$m_b$, and the external momenta; $\Gamma/\gamma_2$ is,
trivially, the identity. Again, in an effective two-Higgs-doublet
Lagrangian, the operator describing $\gamma_2$ is the very operator
$O_7$ in Eq.~(\ref{eq-O7and8}). This gives rise to a vanishing
zeroth-order term in the expansion of light masses, i.e.,  a vanishing
nondecoupling contribution, and was, therefore, not included in the
analyses of Refs.~\cite{NLO-SUSY2a,NLO-SUSY2b}.

The case in which the chirality flip is on the $\tilde{t}$-squark line
instead of on the $t$-quark line can be treated in a similar
way. In this case, the term $\gamma_1$, shown by the lower
right diagram of Fig.~\ref{fig-triansubdiag}, is already of decoupling type
and was also not considered in Refs.~\cite{NLO-SUSY2a,NLO-SUSY2b}.

The procedure to be followed is similar in the case in which the
photon is attached to the $t$-quark line. When the chirality flip is on
the $t$-quark line, the diagram $\gamma_1$ is again the upper right
diagram of Fig.~\ref{fig-triansubdiag}, which at the zeroth-order
expansion gives rise to the same term of the effective 
two-Higgs-doublet Lagrangian given in Eq.~(\ref{eq-2HDMEffLag}). The 
diagram $\Gamma/\gamma_1$ is analogous to the upper left diagram of
Fig.~\ref{fig-triansubdiag}.  The corresponding contributions to the
Wilson coefficient $C_7(\mu_W)$ and to $C_8(\mu_W)$, when the photon
is substituted by a gluon, are
\begin{eqnarray} 
\Delta C_{7,H}^{1\,a}(\mu_W)\vert_{\rm nondec}   
& = &   
\frac{1}{2} \,Q_t \, \tan\beta \, \Delta_{t_R,s} \, 
 \frac{m_t^2}{M_{H^\pm}^2} 
F_3 \!\left(\!\!\frac{m_t^2}{M_{H^\pm}^2}\!\!\right),   
\label{C7tnondec}
\\
\Delta C_{8,H}^{1\,a}(\mu_W)\vert_{\rm nondec}   
& = &
\frac{1}{2} \, \tan\beta \, \Delta_{t_R,s} \, 
 \frac{m_t^2}{M_{H^\pm}^2} 
F_3 \!\left(\!\!\frac{m_t^2}{M_{H^\pm}^2}\!\!\right).
\label{C8tnondec}
\end{eqnarray}
Again, the contributions from the second diagram $\gamma_2$ and the
diagrams obtained when the chirality flip is on the $\tilde{t}$-squark
line are of decoupling type. By comparing 
Eqs.~(\ref{C7Hnondec}),~(\ref{C7tnondec}), and~(\ref{C8tnondec}) with 
the $\tan \beta$-unsuppressed one-loop contributions $C_{i,H}^0(\mu_W)$ 
in Eqs.~(\ref{eq-c70}) and~(\ref{eq-c80}), the relation 
\begin{equation}
\Delta C_{i,H}^1(\mu_W)\vert_{\rm nondec} = 
-\tan\beta \, \Delta_{t_R,s} \, C_{i,H}^0(\mu_W) 
\end{equation}
follows, which shows clearly that $\Delta C_{i,H}^1(\mu_W)$ has the same 
$m_t$ and $m_{H^\pm}$ dependence as that of $C_{i,H}^0(\mu_W)$.

Slightly different is the case in which the photon is emitted by one of
the two squarks $\tilde{t}$,$\tilde{s}$.  In the case of emission from
the $\tilde{s}$ squark, for example, the decomposition of the Feynman
integral has, again, a term corresponding to the full diagram
$F_{\gamma_2}$, to be expanded in $m_t$,$m_{H^\pm}$,$m_b$, and the
external momenta.  The other term is given by the convolution of the
Feynman integral $F_{\gamma_1}$ corresponding to the upper or lower box
diagrams $\gamma_1$ shown on the right of Fig.~\ref{fig-boxphsubdiagr}
(depending on where the chirality is flipped, i.e., on the $t$ quark or
$\tilde{t}$ squark), expanded in the momenta of $H^\pm$, of the $t$ and
the $s$ quarks and the momentum of the photon, and the Feynman integral
$F_{\Gamma/\gamma_1}$ of the upper or lower diagrams ${\Gamma/\gamma_1}$
on the left side of the same figure.

When it is a gluon to be radiated off during the interactions that
lead the $b$ quark into an $s$ quark, there is the additional
possibility of having the gluon emitted from the $\tilde{g}$ line.
The decompositions in $\gamma_1$ and $\Gamma/\gamma_1$ of the
corresponding diagrams are explicitly shown in
Fig.~\ref{fig-boxglsubdiag}.

The results of the calculation in the regime of
Eq.~(\ref{eq-decouplimit}) turn out to be rather involved, already at
$O(m_{\rm weak}^2,m_{H^\pm}^2/M_{\rm SUSY}^2)$ and 
$O((m_{\rm weak}^2,m_{H^\pm}^2/M_{\rm SUSY}^2)^2)$. We report
them only numerically in the next section, together with the numerical
results of the exact calculation of Sec.~\ref{sect-diagrams}.

\section{$H^+$ contribution to $C_{7,8}(\mu_W)$ up to 
 $O(\alpha_s\tan\!\beta)$} 
\label{sect-WCoeff}
We present here numerical results for the charged-Higgs-boson 
contributions to
the Wilson coefficients $C_7$ and $C_8$ at the scale $\mu_W=M_W$, i.e.,
$C_{7,H}(\mu_W)$ and $C_{8,H}(\mu_W)$, including the 
$O(\alpha_s\tan\!\beta)$ corrections discussed in the previous
sections. We make a comparison of the exact results for 
$C_{7,H}(\mu_W)$ and $C_{8,H}(\mu_W)$, which we denote by
$C_{7,H}(\mu_W)\vert_{\rm exact}$ and $C_{8,H}(\mu_W)\vert_{\rm exact}$, 
obtained from Eqs.~(\ref{eq-deltaC7}) and~(\ref{eq-deltaC8}), with 
several approximate forms, $C_{7,H}(\mu_W)\vert_{\rm approx}$ and
$C_{8,H}(\mu_W)\vert_{\rm approx}$: the nondecoupling approximation,
and the two approximations in which the first- and second-order terms
in the expansion of the exact coefficients in 
$(m_{\rm weak}^2, m_{H^{\pm}}^2)/M_{\rm SUSY}^2$ are retained.

\begin{figure}[th] 
\vspace{-1.0truecm}
\begin{center} 
\includegraphics[width= 7.8cm]{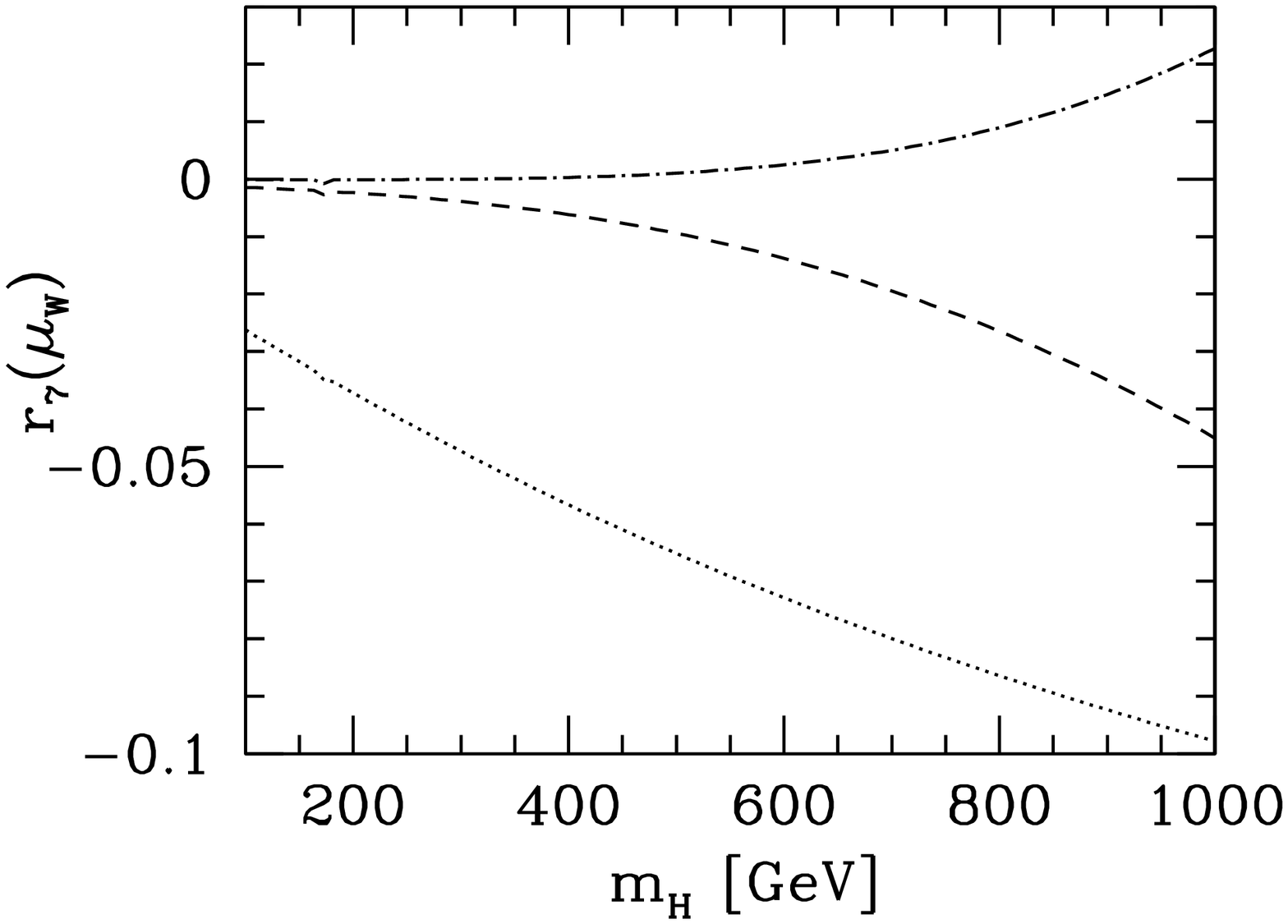}
\hspace{4mm}
\includegraphics[width= 7.8cm]{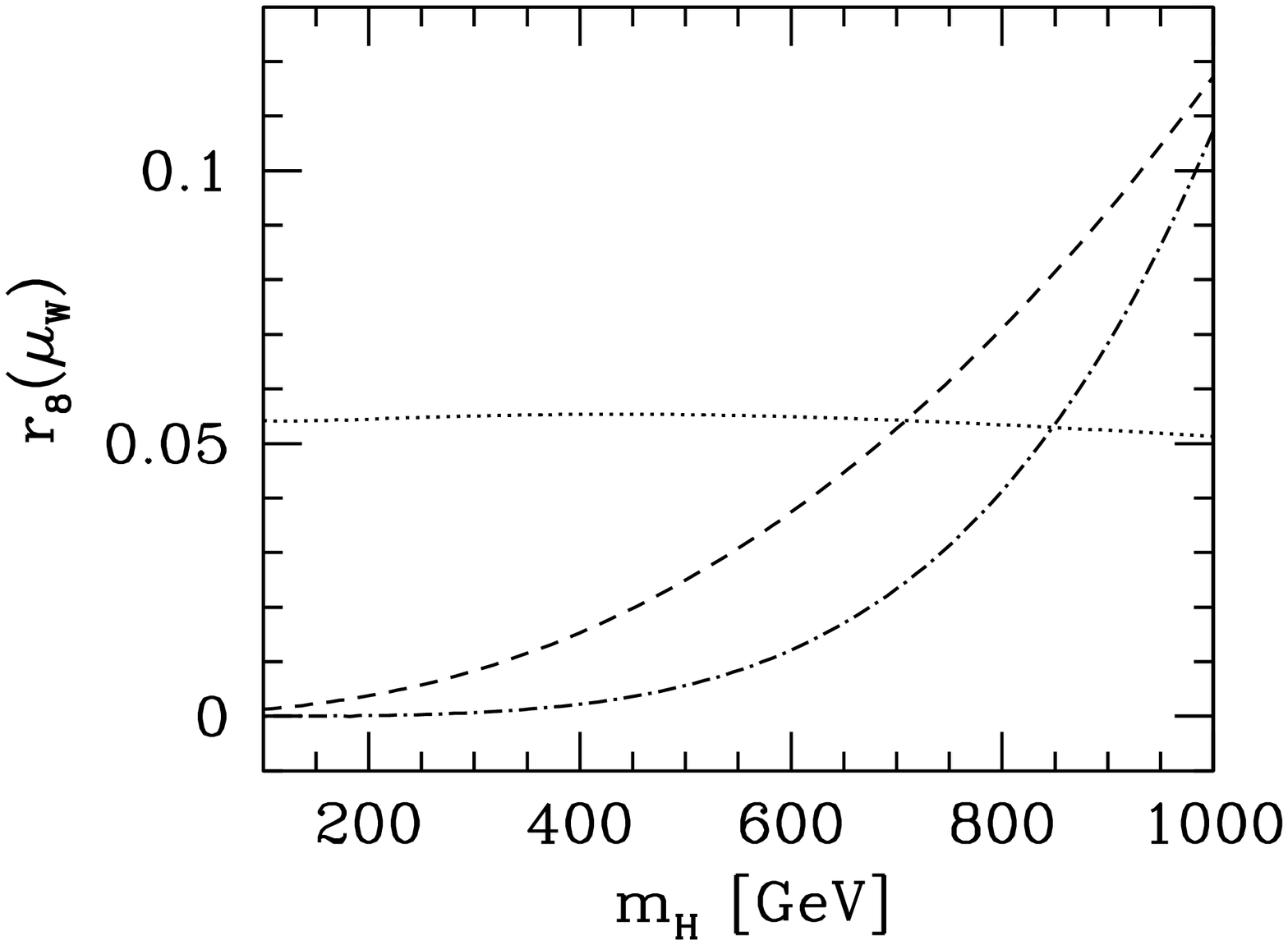} 
\end{center} 
\vspace{-0.3truecm}
\caption[f1]{\small Ratios $r_{7,8}(\mu_W)$ for various approximations
 of the $O(\alpha_s \tan\beta)$ corrections, as defined in the
 text, for spectrum I.  The dotted, dashed, and dot-dashed lines
 show the nondecoupling approximation, first-order expansion, and
 second-order expansion in $(m_t^2,m_{H^\pm}^2)/M_{\rm SUSY}^2$,
 respectively.}
\label{c78figheavy} 
\vspace{-1.0truecm}
\begin{center} 
\includegraphics[width= 7.8cm]{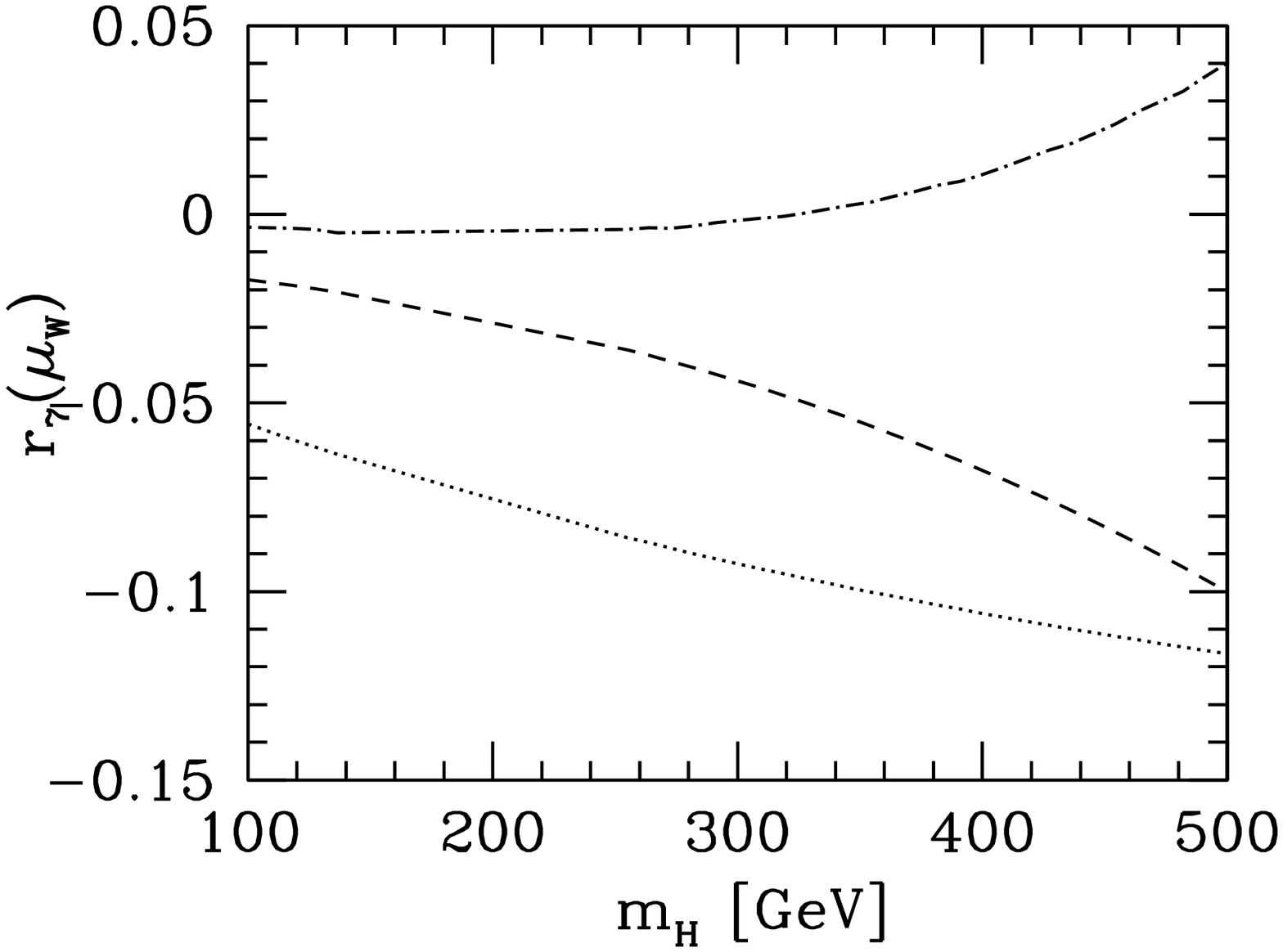}
\hspace{4mm}
\includegraphics[width= 7.8cm]{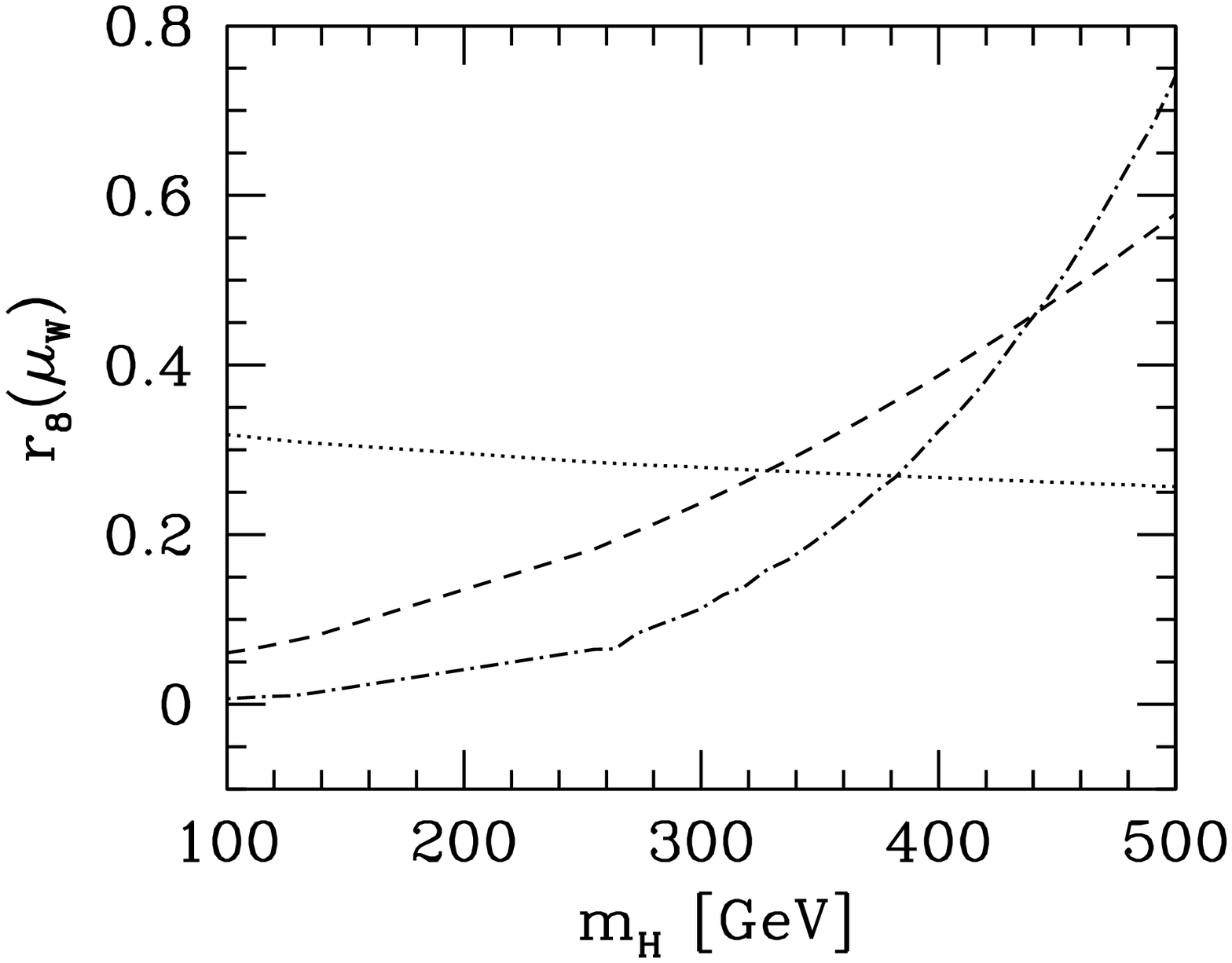} 
\end{center} 
\vspace{-0.3truecm}
\caption[f1]{\small $r_{7,8}(\mu_W)$ for 
 spectrum II. Notation is the same as in Fig.~\ref{c78figheavy}. }
\label{c78figlight} 
\end{figure}

\begin{figure}[th] 
\vspace{-1.0truecm}
\begin{center} 
\includegraphics[width= 7.8cm]{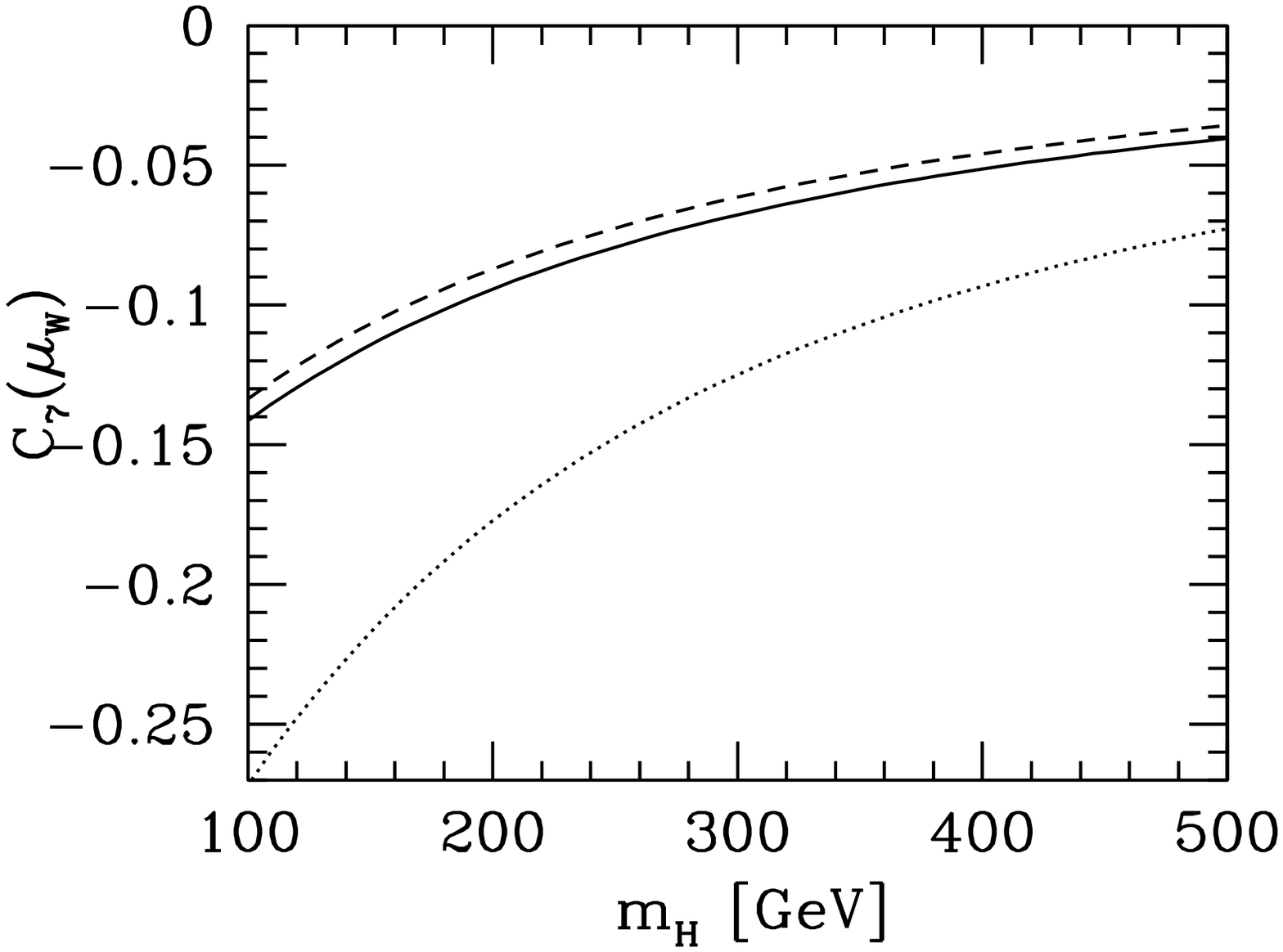}
\hspace{4mm}
\includegraphics[width= 7.8cm]{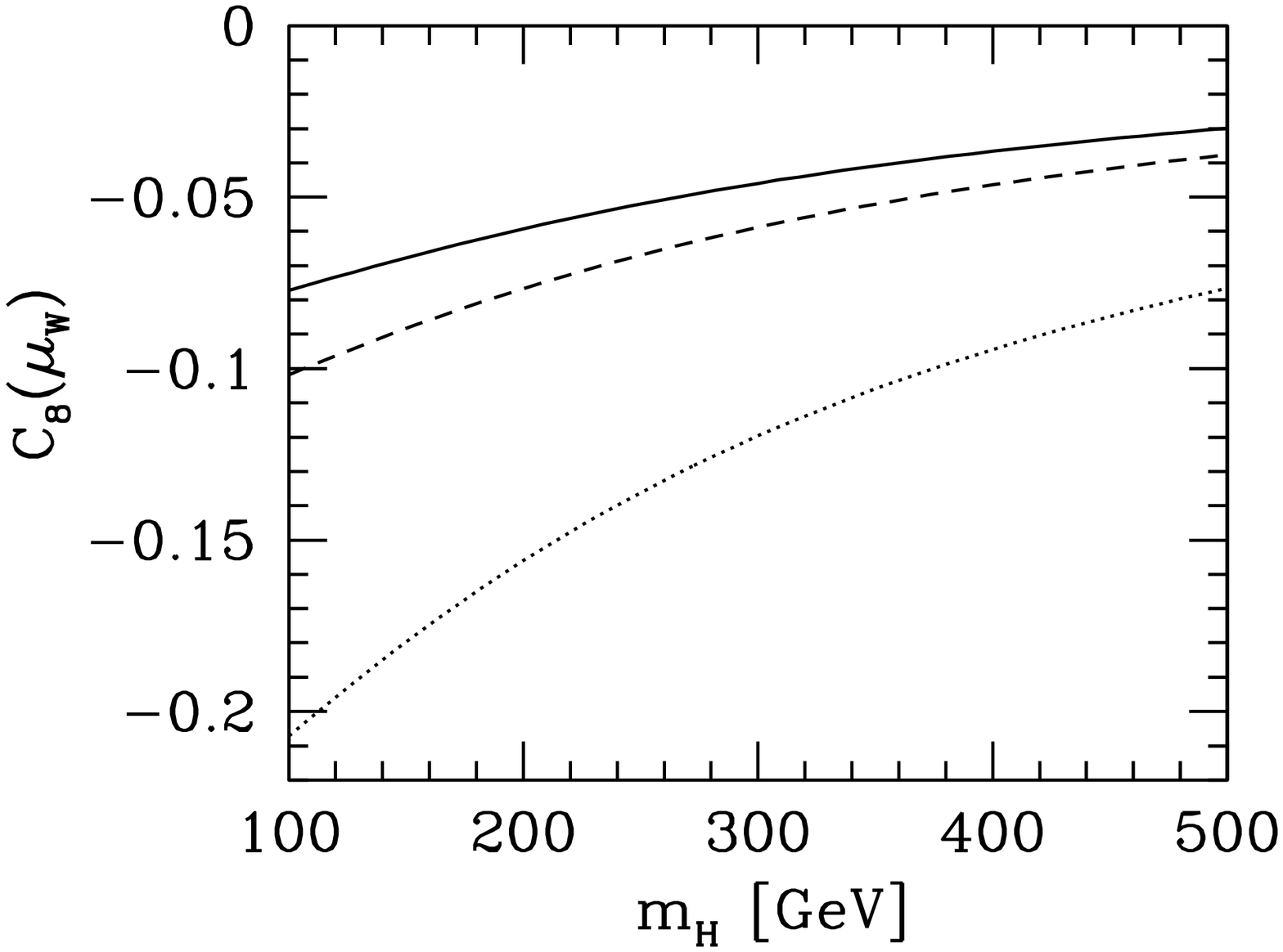} 
\end{center} 
\vspace{-0.3truecm}
\caption[f1]{\small $C_{7,H}(\mu_W)$ and $C_{8,H}(\mu_W)$ for 
 spectrum II.  The dotted, dashed, and solid lines show the one-loop
 result, nondecoupling approximation, and exact two-loop result,
 respectively.}
\label{c78figlight2} 
\end{figure}

In Figs.~\ref{c78figheavy} and~\ref{c78figlight}, we plot the ratios
\begin{equation}
 r_i(\mu_W) \equiv 
\displaystyle{
  \frac{C_{i,H}(\mu_W)\vert_{\rm approx}
       -C_{i,H}(\mu_W)\vert_{\rm exact}}
       {C_{i,H}(\mu_W)\vert_{\rm exact}} } 
\hspace*{5truemm}
 (i = 7,8),
\label{plotratio}
\end{equation}
in which the $O(\alpha_s \tan \beta)$ corrections to the $b$-quark
mass cancel out, showing the goodness of each approximation as a
function of $m_{H^\pm}$. We denote these ratios for the nondecoupling
approximation and those with an expansion in the first and second
order in $(m_t^2, m_{H^\pm}^2)/M_{\rm SUSY}^2$ by dotted, dashed, and
dot-dashed lines, respectively.

Two sets of parameters for supersymmetric particles are used. For
Fig.~\ref{c78figheavy}, we have chosen a superpartner spectrum, called
here spectrum~I, with
$(m_{\tilde{s}_L},m_{\tilde{Q}},m_{\tilde{T}^c},m_{\tilde{B}^c})=
 (700,450,435,470)\,$GeV, $A_t = 150\,$GeV, 
$m_{\tilde{g}}=600\,$GeV, and $\mu=550\,$GeV. 
For Fig.~\ref{c78figlight}, a lighter spectrum is considered:
$(m_{\tilde{s}_L},m_{\tilde{Q}},m_{\tilde{T}^c},m_{\tilde{B}^c})=
 (250,230,210,260)\,$GeV, $A_t = 70\,$GeV, 
$m_{\tilde{g}}=200\,$GeV, and $\mu=250\,$GeV. This is denoted as 
spectrum~II. As for other input parameters, we have used 
$\tan\!\beta=30$, $m_t(\mu_W) = 176.5\,$GeV, which corresponds 
to a pole mass $M_t=175\,$GeV, $m_b(\mu_W)= 3\,$GeV, $A_b=0$, 
$\alpha_s(\mu_W) =0.12 $, $M_Z=91.2$ GeV, and $\sin^2 \theta_W = 0.23$.

For spectrum~I, the difference between the exact calculation and
the nondecoupling approximation is very small in the whole range of
$m_{H^{\pm}}$, even for $m_{H^{\pm}}\gtap M_{\rm SUSY}$.  This is an
unexpected result since, as discussed in Sec.~\ref{sect-EFLvsHME},
the nondecoupling approximation is, in principle, theoretically
justified only for $m_{H^{\pm}}^2\ll M_{\rm SUSY}^2$. In the case of
spectrum~II, $r_{7,8}$ become larger for the nondecoupling
approximation. The corrections beyond this approximation are of the
same order as the SU(2)$\times$U(1) breaking effects in the supersymmetric
particle subloops~\cite{OTHERPAPS} and are no longer negligible.
Nevertheless, $r_7$ and $r_8$ for the nondecoupling approximation
remain of the same order of magnitude for increasing $m_{H^{\pm}}$, 
up to $m_{H^{\pm}}\gg M_{\rm SUSY}$.

In both cases, the bulk of the difference between the results of the
exact and nondecoupling calculations come from $I_{ti2}$ in 
$\Delta C_{7,H}^{1\,a}$ and $\Delta C_{8,H}^{1\,a}$, and also from
$I_{\tilde{g}i2}$ in $\Delta C_{8,H}^{1\,e}$, of 
Eqs.~(\ref{eq-deltaC7}) and~(\ref{eq-deltaC8}).  The inclusion of 
the $(m_t^2,m_{H^\pm}^2/M_{\rm SUSY}^2)^n$ terms by the HME, as 
described in Sec.~\ref{sect-EFLvsHME}, improves the goodness $r_i$ 
of the nondecoupling approximation when $m_{H^\pm}<M_{\rm SUSY}$, but 
worsens it when $m_{H^\pm} \gtap M_{\rm SUSY}$.  This is clearly seen 
in Fig.~\ref{c78figlight}.

In Fig.~\ref{c78figlight2}, we show $C_{7,H}(\mu_W)$ and
$C_{8,H}(\mu_W)$ for the nondecoupling approximation and exact
two-loop results, as well as the one-loop results $C_{i,H}^0(\mu_W)$,
for spectrum~II. The $O(\alpha_s \tan\beta)$ corrections
are comparable to the one-loop contributions.  The deviation of the
exact calculation from the nondecoupling approximation is small, but
nonnegligible for $C_{8,H}$.  It is also shown that all three results
have a similar dependence on $m_{H^\pm}$.

To understand the results for $m_{H^\pm}\gtap M_{\rm SUSY}$
qualitatively, we focus on the diagram in
Fig.~\ref{fig-CHexch2loops}(a), with chirality flip on the $t$-quark
line. When $m_{H^\pm}$ is sufficiently larger than $m_t$, the
contribution of this diagram to $\Delta C_{7,H}^1(\mu_W)$ and 
$\Delta C_{8,H}^1(\mu_W)$ is among the largest.  Analytically, it is
proportional to $\mu m_{\tilde{g}} I_{ti2}$, with the integral
$I_{ti2}$ listed in Eq.~(\ref{eq-intt2}) of
Appendix~\ref{app-integrals}. For the following discussion it is
convenient to write $\mu m_{\tilde{g}} I_{ti2}$ in the form
\begin{equation}
 \mu m_{\tilde{g}} 
 I_{ti2}(m_t,m_{H^\pm}, m_{\tilde{t}_i},m_{\tilde{s}},m_{\tilde{g}}) =
\int\frac{d^4 k}{(2 \pi)^4} \
\frac{k^2}
{\left[k^2 -m_t^2\right]^3 \left[k^2 - m_{H^\pm}^2\right]}
 \,
 Y_{ti2}\left(k^2;m_{\tilde{t}_i},m_{\tilde{s}},m_{\tilde{g}}
       \right),  
\label{eq-Iint}
\end{equation}
where $Y_{ti2}(k^2;m_{\tilde{t}_i},m_{\tilde{s}},m_{\tilde{g}})$
represents the subdiagram contribution to the vertex 
$H^-\bar{s}_Lt_R$ and is given by 
\begin{equation}
Y_{ti2}(k^2;m_{\tilde{t}_i},m_{\tilde{s}},m_{\tilde{g}}) =
\mu m_{\tilde{g}} 
 \left[-2 F + (k^2-m_t^2) G\right]
\left(k^2;m_{\tilde{t}_i}^2,m_{\tilde{s}}^2,m_{\tilde{g}}^2\right),
\label{eq-Ydef} 
\end{equation}
with 
\begin{eqnarray}
F(k^2; m_{\tilde{t}_i}^2,m_{\tilde{s}}^2,m_{\tilde{g}}^2)
 & = & 
\int\frac{d^4 l}{(2 \pi)^4} \
\frac{1}{ \left[ (l+k)^2 - m_{\tilde{t}_i}^2 \right]
\left[ l^2 -\!m_{\tilde{s}}^2 \right]
\left[ l^2-m_{\tilde{g}}^2 \right] },
\label{eq-Fdef}
\\
k^{\mu} 
G(k^2; m_{\tilde{t}_i}^2,m_{\tilde{s}}^2,m_{\tilde{g}}^2) 
 & = & 
\int\frac{d^4 l}{(2 \pi)^4} \
\frac{l^{\mu}}{ \left[ (l+k)^2 - m_{\tilde{t}_i}^2 \right]
\left[ l^2 -\!m_{\tilde{s}}^2 \right]
\left[ l^2-m_{\tilde{g}}^2 \right]^2}.
\label{eq-Gdef}
\end{eqnarray}

The nondecoupling approximation 
$\mu m_{\tilde{g}} I_{ti2}\vert_{\rm nondec}$ of Eq.~(\ref{eq-Iint}),
where the integral $I_{ti2}\vert_{\rm nondec}$ is given in
Eq.~(\ref{eq-ti2-nondec}), is obtained by approximating
$Y_{ti2}(k^2; m_{\tilde{t}_i},m_{\tilde{s}},m_{\tilde{g}})$ by 
\begin{equation}
 Y_{ti2}\vert_{\rm nondec} =
 - 2 \mu m_{\tilde{g}} 
  F (0; m_{\tilde{t}_i}^2,m_{\tilde{s}}^2,m_{\tilde{g}}^2),
\label{eq-Yapprox}
\end{equation}
which is a constant with respect to $k^2$.  Note the relation 
$F (0;m_{\tilde{t}_i}^2,m_{\tilde{s}}^2,m_{\tilde{g}}^2) =
 (-i/16 \pi^2) I(m_{\tilde{t}_i}^2,m_{\tilde{s}}^2,m_{\tilde{g}}^2)$,    
where $I$ is the function defined in Eq.~(\ref{c0function}).  To
simplify our discussion of the behavior of these functions, we 
hereafter set $m_{\tilde{t}_i}$, $m_{\tilde{s}}$, $m_{\tilde{g}}$, and
$\mu$ equal to $M_{\rm SUSY}$.

For $\vert k^2 \vert \ll M_{\rm SUSY}^2$, $F(k^2;M_{\rm SUSY}^2)$ 
and $G(k^2;M_{\rm SUSY}^2)$ behave as 
\begin{eqnarray}
F(k^2; M_{\rm SUSY}^2) 
& = &
 O\left(\frac{1}{M_{\rm SUSY}^{2}}\right) + 
 O\left(\frac{k^2}{M_{\rm SUSY}^{4}}\right), 
\nonumber \\
G(k^2; M_{\rm SUSY}^2) 
& = &
 O\left(\frac{1}{M_{\rm SUSY}^{4}}\right).
\label{eq-FHzeroKlimit}
\end{eqnarray} 
For $\vert k^2 \vert \gg M_{\rm SUSY}^2$, the behavior is
\begin{eqnarray}
F(k^2; M_{\rm SUSY}^2) 
& \to & 
 O\left( \frac{1}{k^2} \ln \frac{k^2}{M^2_{\rm SUSY}} \right), 
\nonumber \\
G(k^2; M_{\rm SUSY}^{2})  
& \to & 
 O\left( \frac{1}{k^4}\ln \frac{k^2}{M_{\rm SUSY}^2} \right).
\end{eqnarray}
The behavior of $Y_{ti2}(k^2; M_{\rm SUSY}^2)$ is therefore 
\begin{equation}
Y_{ti2}(k^2; M_{\rm SUSY}^2)  \to  
\left\{ 
\begin{array}{ll} 
 Y_{ti2}\vert_{\rm nondec} + 
 O\left(
 \displaystyle{\frac{k^2}{M_{\rm SUSY}^2}},
 \displaystyle{\frac{m_t^2}{M_{\rm SUSY}^2}} \right)  
& 
(\vert k^2 \vert \ll M_{\rm SUSY}^2), 
\\[1.8ex] 
 O\left(\displaystyle{\frac{M_{\rm SUSY}^2}{k^2} 
 \ln \frac{k^2}{M_{\rm SUSY}^2}} \right) 
& 
(\vert k^2 \vert \gg M_{\rm SUSY}^2), 
\end{array} 
\right. 
\label{eq-deviation}
\end{equation}
which supports the naive expectation that a substantial deviation of 
$I_{ti2}(m_t,m_{H^\pm},M_{\rm SUSY}^2)$ from 
$I_{ti2}(m_t,m_{H^\pm},M_{\rm SUSY}^2)\vert_{\rm nondec}$ may 
arise from the region $\vert k^2\vert \gtap M_{\rm SUSY}^2$.

The factor multiplying $Y_{ti2}(k^2; M_{\rm SUSY}^{2})$ in
Eqs.~(\ref{eq-Iint}), however, plays a rather important role, leading 
to the fact that this expectation does not hold in the case in which
$M_{\rm SUSY}$ is not rather light.  Since for 
$\vert k^2\vert \gg m_{H^\pm}^2$ this factor drops as $d^4k/k^6$, the
integral~(\ref{eq-Iint}) gets its largest contribution from the region
$\vert k^2\vert \ltap m_{H^\pm}^2$.  A closer inspection actually
shows that it is the region of small $\vert k^2\vert$, up to 
$\vert k^2\vert = O( m_t^2)$, which determines the bulk of the
value of this integral. If $m_t$ is sufficiently smaller than 
$M_{\rm SUSY}$, $Y_{ti2}(k^2; M_{\rm SUSY}^2)$ does not deviate
substantially from $Y_{ti2}\vert_{\rm nondec}$.  The region in $k^2$
where the deviation 
$Y_{ti2}(k^2; M_{\rm SUSY}^2) - Y_{ti2}\vert_{\rm nondec}$ is largest, 
i.e., $\vert k^2\vert \sim M_{\rm SUSY}^2$, is weighted by a rather
efficient suppression factor in Eq.~(\ref{eq-Iint}) if 
$m_t^2 \ll m_{H^\pm}^2 \ltap M_{\rm SUSY}^2 $.

It is clear that an expansion of the integral 
$I_{ti2}(m_t,m_{H^\pm},M_{\rm SUSY}^2)$ in $m_t^2/M_{\rm SUSY}^2$ 
and $m_{H^\pm}^2/M_{\rm SUSY}^2$, as obtained from the HME, generates
terms $O(m_{H^\pm}^0)$ at the first order, and terms 
$O(m_{H^\pm}^2)$ at the second order. These terms contribute to
give a better approximation of the exact results for $C_{i,H}(\mu_W)$,
when $m_{H^\pm}^2 < M_{\rm SUSY}^2$, but have a dependence on
$m_{H^\pm}$ that is rather different from that of the exact results,
when $m_{H^\pm}^2 \gtap M_{\rm SUSY}^2$.  As already observed, the
$m_{H^\pm}$ dependence of the exact results for $C_{i,H}(\mu_W)$ at
order $O(\alpha_s \tan\beta)$ is similar to that of their
nondecoupling approximation, which is, in turn, identical to that of
the one-loop results $C_{i,H}^0(\mu_W)$ (i.e., the results at LO in
QCD) for the same Wilson coefficients.  As for the series expansion
obtained through the HME, the qualitative discussion sketched above
and the limited number of terms we have calculated do not allow us to
conclude whether it is convergent or not.  Even if convergent,
however, the numerical results we have plotted in the previous
figures show clearly that it is not converging fast enough to be of
any practical use in the region $m_{H^\pm}^2 \gtap M_{\rm SUSY}^2$, as
was originally hoped.

\section{Conclusion}
\label{sect-conclu}
In this paper we have evaluated the $O(\alpha_s \tan \beta)$
corrections to the charged-Higgs-boson mediated contributions to the 
Wilson
coefficients relevant for the decay $\bar{B} \to X_s \gamma $, in
supersymmetric models with large $\tan\beta$.  These corrections are
generated by the shift of the $b$-quark mass in the Higgs-boson--quark
couplings and by the dressing of the one-loop 
charged-Higgs-boson diagrams
with squark-gluino subloops (see Fig.~\ref{fig-CHexch2loops}).  The
former corrections are very well known. In this paper we have focused
on the latter class of corrections, which exist in the literature in
an approximate form.

In previous studies~\cite{NLO-SUSY2a,NLO-SUSY2b}, the contributions
from these two-loop diagrams were calculated using an effective
two-Higgs-doublet Lagrangian formalism where the squarks and gluino
are integrated out.  This method is theoretically justified in the
limit in which the charged Higgs boson is light, i.e., at the
electroweak scale $\sim m_{\rm weak}$, whereas the squark and gluino
masses, $ \sim M_{\rm SUSY}$, are rather large.  The resulting
approximation, in which only all the nondecoupling terms in the large
$M_{\rm SUSY}$ limit are retained, has clearly the nontrivial
advantage of providing a rather compact result for these
corrections. However, it is in general expected that its deviation
from the exact two-loop result, of
$O(m_{\rm weak}^2,m_{H^\pm}^2/M_{\rm SUSY}^2)$, becomes 
significant for $m_{\rm weak} \sim M_{\rm SUSY} $ and/or 
$m_{H^\pm} \ge M_{\rm SUSY}$.

We have calculated the contributions of the two-loop diagrams in
Fig.~\ref{fig-CHexch2loops} exactly, without assuming any patterns for
the masses of the particle involved.  By making use of the heavy mass
expansion technique, we have also evaluated two additional
approximate forms for the same diagrams, including all terms up to
$O(m_{\rm weak}^2,m_{H^\pm}^2/M_{\rm SUSY}^2)$ in one case, and
all terms up to
$O((m_{\rm weak}^2,m_{H^\pm}^2/M_{\rm SUSY}^2)^2)$ in the
other. The exact calculation, not confined to specific values of the
masses involved, has allowed us to establish the goodness of all three
approximations.

Our findings can be summarized as follows.  Surprisingly, the results
of Refs.~\cite{NLO-SUSY2a,NLO-SUSY2b} approximate the exact two-loop
result quite adequately, irrespective of the value of $m_{H^\pm}$
relative to $M_{\rm SUSY}$, provided $M_{\rm SUSY}$ is large
enough. The unexpected absence of a large deviation for the case of
$m_{H^\pm} \gtap M_{\rm SUSY}$ with $m_{\rm weak}^2\ll M_{\rm SUSY}^2$ 
can be understood from the structure of the two-loop integrals.  This
points to the fact that the only relevant hierarchy in this problem is
that between $m_{\rm weak}$ and $M_{\rm SUSY}$, whereas, for
$m_{H^\pm} > m_{\rm weak}$, the value of $m_{H^\pm}$ with respect to
that of $M_{\rm SUSY}$ plays a rather marginal role.  Therefore,
deviations between the exact result and that of the nondecoupling
approximation can be found only for $M_{\rm SUSY}$ not much larger
than $m_{\rm weak}$.  The inclusion of the higher-order terms
$O(m_{\rm weak}^2,m_{H^\pm}^2/M_{\rm SUSY}^2)$ and 
$O((m_{\rm weak}^2,m_{H^\pm}^2/M_{\rm SUSY}^2)^2)$
improves the approximation of Refs.~\cite{NLO-SUSY2a,NLO-SUSY2b} 
for $m_{H^\pm}^2 \ll M_{\rm SUSY}^2$, as expected, but makes it worse for
$m_{H^\pm} \gtap M_{\rm SUSY}$.  This behavior is attributed to the
fact that these higher-order terms in the 
$(m_{\rm weak}^2,m_{H^\pm}^2/M_{\rm SUSY}^2)$ expansion have a
dependence on $m_{H^\pm}$ rather different from that of the lowest-order
terms in this expansion (the terms of the nondecoupling approximation) 
and that of the exact two-loop result.

We have illustrated our findings by showing the values of the Wilson
coefficients $C_7$ and $C_8$ at the electroweak matching scale
$\mu_W$, for different gluino, squark, and charged Higgs boson 
spectra.  We
postpone a presentation of the same coefficients $C_7$ and $C_8$ at a
low scale $\sim m_b$ and of the actual branching ratio 
${\rm BR}(\bar{B} \to X_s \gamma)$ to future work~\cite{BGYfuture}.

\vspace*{0.5truecm}
\noindent 
{\bf Acknowledgments}  
The authors acknowledge the Theory Groups of the Universitat
Aut\`onoma de Barcelona, the University of Bern, KEK, Tohoku
University, the University of Tokyo, and the Yokohama National
University for hospitality at different stages of this work.
F.B. was supported in part by the Japanese Society for Promotion of
Science; C.G. by the Schweizerischer Nationalfonds
and by RTN, BBW Contract No.~01.0357, and EC Contract 
No.~HPRN-CT-2002-00311 (EURIDICE); Y.Y. by the Grant-in-Aid for 
Scientific Research from the Ministry of Education, Culture, Sports, 
Science, and Technology of Japan, No.~14740144.

\appendix
\section{}
\label{app-funct}
We report here the functions $F_3(x)$ and $F_4(x)$ introduced in 
the second paper of Ref.~\cite{BSGcontributions}:
\begin{eqnarray}
 F_3(x) & = &
  \frac{1}{ 2 (x-1)^3} \left( x^2 -4x +3 +2\log x\right),
\nonumber \\
 F_4(x) & = &
  \frac{1}{ 2 (x-1)^3} \left( x^2 -1 -2x\log x\right).
\label{app-f3and4}
\end{eqnarray}

\section{}
\label{app-conventions}
We have adopted the following conventions for squark mass 
eigenstates $\tilde{q}_1$,$\tilde{q}_2$ and mass 
eigenvalues $m_{\tilde{q}_1}^2$,$ m_{\tilde{q}_2}^2$:  
\begin{equation}
 \left(
\begin{array}{c}  \tilde{q}_L \\ \tilde{q}_R \end{array}
 \right)  =  U_q 
 \left(
\begin{array}{c}  \tilde{q}_1 \\ \tilde{q}_2 \end{array}
 \right),
\label{sfdiag}
\end{equation}
where the diagonalization matrix $ U_q$ is such that
\begin{equation}
 U_q^T {\cal M}^2_{\tilde{q}} \,U_q 
\equiv 
 {\widehat{\cal M}}^2_{\tilde{q}} 
= 
\left(
\begin{array}{cc}
 m_{\tilde{q}_1}^2    &   0   \\
      0               &  m_{\tilde{q}_2}^2 
\end{array}
\right).
\label{massdiag}
\end{equation}
${\cal M}^2_{\tilde{q}}$ is the mass squared matrix for the 
squark $\tilde{q}$ and ${\widehat{\cal M}}^2_{\tilde{q}}$ is the 
diagonalized squark mass squared. For $\tilde{q}=\tilde{b}$ and $\tilde{t}$, 
the matrix ${\cal M}^2_{\tilde{q}}$ is, respectively,
\begin{equation}
{\cal M}^2_{\tilde{b}} = 
\left(
\begin{array}{cc}
 m_{\tilde{Q}}^2 + m_b^2 + D_L^b   & 
 m_b(A_b - \mu \tan \beta) 
\\
 m_b(A_b - \mu \tan \beta)         & 
 m_{\tilde{B}^c}^2 + m_b^2 + D_R^b 
\end{array}
\right) 
\label{sbotmm} 
\end{equation}
and
\begin{equation}
{\cal M}^2_{\tilde{t}} = 
\left(
\begin{array}{cc}
 m_{\tilde{Q}}^2 + m_t^2 + D_L^t   & 
 m_t(A_t - \mu \cot \beta) 
\\
 m_t(A_t - \mu \cot \beta)         & 
 m_{\tilde{T}^c}^2 + m_t^2 + D_R^t 
\end{array}
\right),
\label{stopmm}
\end{equation}
with $ D_{L,R}^t$ and $ D_{L,R}^b$ given by
\begin{eqnarray}
 D_L^t  \ = \   \cos 2 \beta M_Z^2 
  \,\left(+\frac{1}{2} - \frac{2}{3} \sin^2\theta_W\right),
 & \quad \quad & 
 D_R^t  \ = \   \cos 2 \beta M_Z^2 
  \,\left(+\frac{2}{3} \sin^2\theta_W\right),
\nonumber \\
 D_L^b  \ = \   \cos 2 \beta M_Z^2 
  \,\left(-\frac{1}{2} + \frac{1}{3} \sin^2\theta_W\right),
 &             & 
 D_R^b  \ = \   \cos 2 \beta M_Z^2 
  \,\left( -\frac{1}{3} \sin^2\theta_W\right).
\end{eqnarray}

\section{}
\label{app-integrals}
\noindent 
After defining the following product of propagators:
\begin{equation}
 D
\equiv
\frac{1}{\left[ k^2    -m_t^2            \right]  
         \left[ k^2    -m_{H^\pm}^2            \right]
         \left[ (l+k)^2-m_{\tilde{t}_i}^2\right]
         \left[ l^2    -m_{\tilde{s}}^2  \right] 
         \left[ l^2    -m_{\tilde{g}}^2  \right]},
\label{eq-propags}
\end{equation}
the integrals in Eqs.~(\ref{eq-deltaC7}) and~(\ref{eq-deltaC8}) 
can be cast in the form
\begin{eqnarray}
I_{ti1}  
& = & 
\int\frac{d^4 l}{(2 \pi)^4} \
\int\frac{d^4 k}{(2 \pi)^4} \
\frac{D}{\left[ k^2 -m_t^2 \right]}
\left\{ \frac{ 2 m_t^2 (l\cdot k)  }{k^2 -m_t^2}   
       -\frac{1}{3} 
        \frac{4(l\cdot k)^2-l^2k^2 }{l^2 -m_{\tilde{g}}^2} \right\},
\label{eq-intt1}
\\[1.5ex]
 I_{ti2}  
& = &
\int\frac{d^4 k}{(2 \pi)^4} \
\int\frac{d^4 l}{(2 \pi)^4} \
\frac{D}{\left[ k^2 -m_t^2 \right]}
\left\{ \frac{l\cdot k}{l^2 -m_{\tilde{g}}^2} 
       -\frac{2k^2}{k^2 -m_t^2} \right\}
\label{eq-intt2}
\end{eqnarray}
for the case in which a photon/gluon is emitted by the $t$ quark; 
\begin{eqnarray}
I_{Hi1} 
& = & 
\int\frac{d^4 k}{(2 \pi)^4} \
\int\frac{d^4 l}{(2 \pi)^4} \
\frac{D}{\left[ k^2    -m_{H^\pm}^2            \right]}
\left\{ \frac{1}{3}
        \frac{4(l\cdot k)^2 -l^2k^2}{l^2 -m_{\tilde{g}}^2} 
       -\frac{ m_t^2 (l\cdot k)}{ k^2 -m_t^2} \right\},
\label{eq-intH1} 
\\[1.5ex]
I_{Hi2} 
& = & 
\int\frac{d^4 k}{(2 \pi)^4} \
\int\frac{d^4 l}{(2 \pi)^4} \
\frac{D}{\left[ k^2    -m_{H^\pm}^2          \right]}
\left\{ \frac{k^2}{ k^2 -m_t^2}
       -\frac{l\cdot k}{l^2 -m_{\tilde{g}}^2} \right\} 
\label{eq-intH2} 
\end{eqnarray} 
for the case of a photon emitted by the charged Higgs boson; 
\begin{eqnarray}
I_{\tilde{t}i1}
& = & 
\int\frac{d^4 k}{(2 \pi)^4} \
\int\frac{d^4 l}{(2 \pi)^4} \
\frac{D}{\left[ (l+k)^2 -m_{\tilde{t}_i}^2\right]} \
\frac{1}{3}
\left\{ \frac{4(l\cdot k)^2-l^2k^2+3 k^2(l\cdot k)}{k^2-m_t^2}
  -\frac{ 4(l\cdot k)^2-l^2k^2+3 l^2(l\cdot k) }
        { l^2-m_{\tilde{g}}^2 } \right\},
\label{eq-intst1}
\\[1.6ex]
I_{\tilde{t}i2}
& = & 
\int\frac{d^4 k}{(2 \pi)^4} \
\int\frac{d^4 l}{(2 \pi)^4} \
\frac{D}{\left[ (l+k)^2 -m_{\tilde{t}_i}^2\right]}
\left\{ \frac{l\cdot(l+k)}{ l^2-m_{\tilde{g}}^2 }
       -\frac{k\cdot(l+k)}{k^2-m_t^2}   \right\}
\label{eq-intst2}
\end{eqnarray}
for the case of a photon/gluon emitted by the $\tilde{t}$ squark;
\begin{eqnarray}
I_{\tilde{s}i1}
& = & 
\int\frac{d^4 k}{(2 \pi)^4} \
\int\frac{d^4 l}{(2 \pi)^4} \
\frac{D}{\left[ l^2     -m_{\tilde{s}}^2  \right]}
\left\{ \frac{1}{3}\frac{4(l\cdot k)^2-l^2k^2 }{k^2-m_t^2} 
       -\frac{m_{\tilde{g}}^2(l\cdot k)}
             { l^2-m_{\tilde{g}}^2 } \right\},
\label{eq-intss1}
\\[1.5ex]
I_{\tilde{s}i2}
& = & 
\int\frac{d^4 k}{(2 \pi)^4} \
\int\frac{d^4 l}{(2 \pi)^4} \
\frac{D}{\left[ l^2     -m_{\tilde{s}}^2  \right]}
\left\{ \frac{l^2}{ l^2-m_{\tilde{g}}^2 }
       -\frac{l\cdot k}{k^2-m_t^2} \right\}
\label{eq-intss2}
\end{eqnarray}
for the case of a photon/gluon emitted by the $\tilde{s}$ squark;
\begin{eqnarray}
I_{\tilde{g}1}
& = & 
\frac{1}{2}
\int\frac{d^4 k}{(2 \pi)^4} \
\int\frac{d^4 l}{(2 \pi)^4} \
\frac{D}{\left[ l^2     -m_{\tilde{g}}^2  \right]}
\left\{ \frac{2m_{\tilde{g}}^2(l\cdot k)}{l^2-m_{\tilde{g}}^2 }
       -\frac{1}{3} 
        \frac{4(l\cdot k)^2-l^2k^2}{k^2-m_t^2}\right\},
\label{eq-intgl1}
\\[1.5ex]
I_{\tilde{g}2}
& = & 
\frac{1}{2}
\int\frac{d^4 k}{(2 \pi)^4} \
\int\frac{d^4 l}{(2 \pi)^4} \
\frac{D}{\left[ l^2     -m_{\tilde{g}}^2  \right]}
\left\{ \frac{l\cdot k}{k^2-m_t^2} 
       -\frac{2l^2}{ l^2-m_{\tilde{g}}^2 }\right\} 
\label{eq-intgl2}
\end{eqnarray}
for the case of a gluon emitted by the gluino $\tilde{g}$.

These integrals were obtained from those corresponding to the diagrams
(a)--(e) in Fig.~\ref{fig-CHexch2loops} after an expansion
on the external momenta and a reduction of all tensorial structures to
scalar ones.  They can be further reduced to linear combinations of
the two-loop vacuum integrals
\begin{eqnarray}
G(m_1,p_1; m_2;p_2;m_3,p_3) 
& = & 
\int\frac{d^d k}{(2 \pi)^d} \ 
\int\frac{d^d l}{(2 \pi)^d} \
\frac{1}{\left[(l+k)^2-m_1^2 \right]^{p_1} 
         \left[l^2    -m_2^2 \right]^{p_2} 
         \left[ k^2   -m_3^2 \right]^{p_3} },
\nonumber \\
&   & 
\label{eq-masterint}
\end{eqnarray}
whose general solution is explicitly reported in Ref.~\cite{GvdB}.

In the nondecoupling approximation of
Refs.~\cite{NLO-SUSY2a,NLO-SUSY2b}, only the integrals $I_{ti2}$ and
$I_{Hi2}$ are needed. They are evaluated after substituting $(l+k)^2$
with $l^2$ in the expression for $D$.  The terms proportional to
$l\cdot k$ in the curly brackets of Eqs.~(\ref{eq-intt2})
and~(\ref{eq-intH2}) are then dropped and the integrals are factorized
into two one-loop integrals as
\begin{eqnarray}
 I_{ti2}\vert_{\rm nondec}
& = &
\int\frac{d^4 k}{(2 \pi)^4} \
\frac{-2k^2}{\left[ k^2 -m_t^2 \right]^3
         \left[ k^2    -m_{H^\pm}^2     \right] }
\int\frac{d^4 l}{(2 \pi)^4} \
\frac{1}{
         \left[ l^2-m_{\tilde{t}_i}^2\right]
         \left[ l^2    -m_{\tilde{s}}^2  \right]
         \left[ l^2    -m_{\tilde{g}}^2  \right]},
\label{eq-ti2-nondec}
\\[1.5ex]
I_{Hi2}\vert_{\rm nondec}
& = &
\int\frac{d^4 k}{(2 \pi)^4} \
\frac{k^2}{\left[ k^2    -m_t^2          \right]^2
           \left[ k^2    -m_{H^\pm}^2    \right]^2 }
\int\frac{d^4 l}{(2 \pi)^4} \
\frac{1}{
         \left[ l^2-m_{\tilde{t}_i}^2\right]
         \left[ l^2    -m_{\tilde{s}}^2  \right]
         \left[ l^2    -m_{\tilde{g}}^2  \right]}.
\label{eq-Hi2-nondec}
\end{eqnarray}


\begin{thebibliography}{99}

\bibitem{NLOSM}
The importance of the QCD corrections for the decay 
$\bar{B} \to X_s \gamma$ was first recognized by 
%
S.~Bertolini, F.~Borzumati and A.~Masiero,
Phys.\ Rev.\ Lett.\  {\bf 59}, 180 (1987);
%
N.~G.~Deshpande, P.~Lo, J.~Trampetic, G.~Eilam and P.~Singer,
{\it ibid.} {\bf 59}, 183 (1987).
%
A summary of the different steps that brought to the completion of the
NLO calculation, after this realization, can be found in
%
A.~J.~Buras and M.~Misiak,
Acta Phys.\ Pol.\ B {\bf 33}, 2597 (2002).
%
See also 
P.~Gambino, M.~Gorbahn and U.~Haisch,
Nucl.\ Phys.\ {\bf B673}, 238 (2003),
for a final test of the NLO calculation not included 
in the previous summary.


\bibitem{EXPERIMENTS} 
Belle Collaboration, K.~Abe {\it et al.}, 
Phys.\ Lett.\ B {\bf 511}, 151 (2001); 
%
CLEO Collaboration, S.~Chen {\it et al.}, 
Phys.\ Rev.\ Lett.\  {\bf 87}, 251807 (2001);
%
BABAR Collaboration, B.~Aubert {\it et al.}, 
hep-ex/0207074; 
hep-ex/0207076. 


\bibitem{NNLOproject}
K.~Bieri, C.~Greub and M.~Steinhauser,
Phys.\ Rev.\ D {\bf 67}, 114019 (2003);
%
M.~Misiak, talk at the 2003 Ringberg Phenomenology Workshop 
 on Heavy Flavours. 


\bibitem{CRS}
P.~Ciafaloni, A.~Romanino and A.~Strumia,
Nucl.\ Phys.\ {\bf B524}, 361 (1998).


\bibitem{NLO2HDM-CDGG}
M.~Ciuchini, G.~Degrassi, P.~Gambino and G.~F.~Giudice,
Nucl.\ Phys.\ {\bf B527}, 21 (1998).


\bibitem{NLO2HDM-BG}
F.~M.~Borzumati and C.~Greub,
Phys.\ Rev.\ D {\bf 58} ,074004 (1998),
%
Phys.\ Rev.\ D {\bf 59}, 057501 (1999);
%
and 
hep-ph/9810240.


\bibitem{BSGinSUSYproposal}
S.~Bertolini, F.~Borzumati and A.~Masiero,
Phys.\ Lett.\ B {\bf 192}, 437 (1987).


\bibitem{LO-generalSUSY}
F.~Borzumati, C.~Greub, T.~Hurth and D.~Wyler,
Phys.\ Rev.\ D {\bf 62}, 075005 (2000);
%
C.~Greub, T.~Hurth and D.~Wyler,
hep-ph/9912420;
%
D.~Wyler and F.~Borzumati,
hep-ph/0104046;


\bibitem{NLO-SUSY1}
M.~Ciuchini, G.~Degrassi, P.~Gambino and G.~F.~Giudice,
Nucl.\ Phys.\ {\bf B534}, 3 (1998);


\bibitem{NLO-SUSY2a}
G.~Degrassi, P.~Gambino and G.~F.~Giudice,
J. High Energy Phys. {\bf 12}, 009 (2000). 


\bibitem{NLO-SUSY2b}
M.~Carena, D.~Garcia, U.~Nierste and C.~E.~M.~Wagner,
Phys.\ Lett.\ B {\bf 499}, 141 (2001). 


\bibitem{MFVBuras}
For a recent discussion on this topic, see
A.~J.~Buras,
Acta Phys.\ Polon.\ B {\bf 34}, 5615 (2003).


\bibitem{largeTB}
H.~Georgi,
in {\it Particles and Fields}--1974, 
edited by C. E. Carlson, AIP Conf. Proc. No. 23 
(AIP, New York, 1975) p. 575; 
%
H.~Fritzsch and P.~Minkowski,
Ann. Phys. (N.Y.) {\bf 93}, 193 (1975);
%
H.~Georgi and D.~V.~Nanopoulos,
Nucl.\ Phys.\ {\bf B155}, 52 (1979);
%
Nucl.\ Phys.\ {\bf B159}, 16 (1979);
%
L.~E.~Ib\'a\~nez and G.~G.~Ross,
Phys.\ Lett.\ B {\bf 105}, 439 (1981);
%
M.~Olechowski and S.~Pokorski,
Phys.\ Lett.\ B {\bf 214}, 393 (1988).


\bibitem{bsglargeTB}
N.~Oshimo,
Nucl.\ Phys.\ {\bf B404}, 20 (1993);
%
F.~M.~Borzumati,
Z.\ Phys.\ C {\bf 63}, 291 (1994);
%
F.~M.~Borzumati, M.~Olechowski and S.~Pokorski,
Phys.\ Lett.\ B {\bf 349}, 311 (1995);
%
H.~Murayama, M.~Olechowski and S.~Pokorski,
Phys.\ Lett.\ B {\bf 371}, 57 (1996);
%
R.~Rattazzi and U.~Sarid,
Nucl.\ Phys.\ {\bf B501}, 297 (1997); 
%
F.~M.~Borzumati,
arXiv:hep-ph/9702307;
%
T.~Bla\v{z}ek and S.~Raby,
Phys.\ Rev.\ D {\bf 59}, 095002 (1999);
%
D.~A.~Demir and K.~A.~Olive,
Phys.\ Rev.\ D {\bf 65}, 034007 (2002).


\bibitem{dmb}
T.~Banks,
Nucl.\ Phys.\ {\bf B303}, 172 (1988);
%
R.~Hempfling,
Phys.\ Rev.\ D {\bf 49}, 6168 (1994);
%
L.~J.~Hall, R.~Rattazzi and U.~Sarid,
Phys.\ Rev.\ D {\bf 50}, 7048 (1994);
%
M.~Carena, M.~Olechowski, S.~Pokorski and C.~E.~M.~Wagner,
Nucl.\ Phys.\ {\bf B426}, 269 (1994);
%
T.~Bla\v{z}ek, S.~Raby and S.~Pokorski,
Phys.\ Rev.\ D {\bf 52}, 4151 (1995);
%
D.~M.~Pierce, J.~A.~Bagger, K.~T.~Matchev and R.~j.~Zhang,
Nucl.\ Phys.\ {\bf B491}, 3 (1997);
%
F.~Borzumati, G.~R.~Farrar, N.~Polonsky and S.~Thomas,
{\it ibid.}\ Phys.\ {\bf B555}, 53 (1999).


\bibitem{carenaH0}
M.~Carena, S.~Mrenna and C.~E.~M.~Wagner,
Phys.\ Rev.\ D {\bf 60}, 075010 (1999);
%
K.~S.~Babu and C.~F.~Kolda,
Phys.\ Lett.\ B {\bf 451}, 77 (1999);
%
F.~Borzumati, G.~R.~Farrar, N.~Polonsky and S.~Thomas,
in Ref.~\cite{dmb};
%
H.~Eberl, K.~Hidaka, S.~Kraml, W.~Majerotto and Y.~Yamada,
Phys.\ Rev.\ D {\bf 62}, 055006 (2000); 
%
M.~Carena, D.~Garcia, U.~Nierste and C.~E.~M.~Wagner,
Nucl.\ Phys.\ {\bf B577}, 88 (2000); 
%
H.~E.~Haber, M.~J.~Herrero, H.~E.~Logan, S.~Pe\~naranda, 
S.~Rigolin and D.~Temes,
Phys.\ Rev.\ D {\bf 63}, 055004 (2001);
%
H.~E.~Logan,
Nucl.\ Phys.\ Proc.\ Suppl.\ {\bf 101}, 279 (2001);
%
M.~J.~Herrero, S.~Pe\~naranda and D.~Temes,
Phys.\ Rev.\ D {\bf 64}, 115003 (2001).


\bibitem{DIAZ}
M.~A.~D\'{\i}az,
Phys.\ Lett.\ B {\bf 304}, 278 (1993).


\bibitem{OTHERPAPS}
G.~D'Ambrosio, G.~F.~Giudice, G.~Isidori and A.~Strumia,
Nucl.\ Phys.\ {\bf B645}, 155 (2002); 
%
A.~J.~Buras, P.~H.~Chankowski, J.~Rosiek and \L.~S{\l}awianowska,
{\it ibid.}\ {\bf B659}, 3 (2003).


\bibitem{BGY} 
F.~Borzumati, C.~Greub and Y.~Yamada,
hep-ph/0305063.


\bibitem{HME} 
V.~A.~Smirnov,
{\it ``Renormalization and Asymptotic Expansions,''}
Progress in physics Vol.~14 
(Birkhaeuser, Basel, Switzerland, 1991); 
%
Mod. Phys. Lett. A {\bf 10}, 1485 (1995);
%
{\it ``Applied Asymptotic Expansions In Momenta And Masses,''}
Springer Tracts in Modern Physics Vol.~177 
(Springer, Berlin, 2002). 


\bibitem{BSGcontributions}
S.~Bertolini, F.~Borzumati and A.~Masiero,
Nucl.\ Phys.\ {\bf B294}, 321 (1987);
%
S.~Bertolini, F.~Borzumati, A.~Masiero and G.~Ridolfi,
{\it ibid.}\ {\bf B353}, 591 (1991).


\bibitem{BGYfuture}
F.~Borzumati, C.~Greub and Y.~Yamada, to appear. 


\bibitem{GvdB}
A.~I.~Davydychev and J.~B.~Tausk,
Nucl.\ Phys.\ {\bf B397}, 123 (1993).
%
A.~Ghinculov and J.~J.~van der Bij,
{\it ibid.}\ {\bf B436}, 30 (1995).


\end{thebibliography}
\end{document}